\documentclass[prl,aps,twocolumn,floatfix,amsmath,amssymb,superscriptaddress]{revtex4-1}
\usepackage{latexsym}
\usepackage{graphicx}
\usepackage{tabularx}
\usepackage{times}
\usepackage[usenames,dvipsnames]{color}
\usepackage{amsmath}
\usepackage{dcolumn}
\usepackage{latexsym,amsmath,amssymb,bm,euscript}
\usepackage[
	colorlinks=true,
	urlcolor=BlueViolet,
 citecolor=Maroon,
	plainpages=false,
 	pdfpagelabels,
 	bookmarksnumbered
 	]{hyperref}
\newcommand{\sgn}{\textrm{sgn}}

\newcommand{\br}{{\bm r}}
\newcommand{\bq}{{\bm q}}
\newcommand{\cK}{{K}}

\renewcommand{\(}{\left(}
\renewcommand{\)}{\right)}

\definecolor{gray}{rgb}{0.4,.4,0.4}
\definecolor{purple}{rgb}{0.6,.0,0.6}
\definecolor{darkgreen}{rgb}{0.0,.6,0.}

\def\beq{\begin{equation}}
\def\eeq{\end{equation}}

\begin{document}

\title{Explicit derivation of duality between a free Dirac cone and \\ quantum electrodynamics in $(2+1)$ dimensions}

\author{David F. Mross}
\affiliation{Department of Physics and Institute for Quantum Information and Matter, California Institute of Technology, Pasadena, CA 91125, USA}

\author{Jason Alicea}
\affiliation{Department of Physics and Institute for Quantum Information and Matter, California Institute of Technology, Pasadena, CA 91125, USA}
\affiliation{Walter Burke Institute for Theoretical Physics, California Institute of Technology, Pasadena, CA 91125, USA}

\author{Olexei I. Motrunich}
\affiliation{Department of Physics and Institute for Quantum Information and Matter, California Institute of Technology, Pasadena, CA 91125, USA}
\affiliation{Walter Burke Institute for Theoretical Physics, California Institute of Technology, Pasadena, CA 91125, USA}

\begin{abstract}
We explicitly derive the duality between a free electronic Dirac cone and quantum electrodynamics in $(2+1)$ dimensions (QED$_3$) with $N = 1$ fermion flavors. The duality proceeds via an exact, non-local mapping from electrons to dual fermions with long-range interactions encoded by an emergent gauge field. This mapping allows us to construct parent Hamiltonians for exotic topological-insulator surface phases, derive the particle-hole-symmetric field theory of a half-filled Landau level, and nontrivially constrain QED$_3$ scaling dimensions. We similarly establish duality between bosonic topological insulator surfaces and $N = 2$ QED$_3$. 
\end{abstract}

\maketitle

{\bf \emph{Introduction.}}~Recent theoretical work has revealed intricate connections between the following rather different physical systems: metallic surfaces of 3D topological insulators (TI's) \cite{FuTI,MooreTI,RoyTI,KaneReview,QiReview}; composite Fermi liquids (CFL's) \cite{HLR,WillettCFL,KangCFL,GoldmanCFL,DasSarmaBook,Willet,JainBook} that arise in a strong magnetic field when the lowest Landau level is half filled; and quantum electrodynamics in $(2+1)$ dimensions (QED$_3$). Central to these surprising relationships is a newly discovered duality between a single Dirac cone with action
\begin{align}
{\cal S}_\text{Dirac} &= \int_{t, x, y} i \bar{{\Psi}} \gamma^\mu (\partial_\mu - i A_\mu)\Psi\label{eq.freedirac},
\end{align}
and QED$_3$ with $N = 1$ fermion flavors,
\begin{align}
{\cal S}_\text{{QED}$_3$} &= \int_{t, x, y} \left[i \bar{\tilde {\Psi}} \gamma^\mu (\partial_\mu - i a_\mu)\tilde \Psi + \frac{\epsilon_{\mu\nu\kappa}A_\mu \partial_\nu a_\kappa}{4\pi} +\cdots\right] \label{eqn.qed3}.
\end{align}
Here $\gamma^\mu$ are $2 \times 2$ Dirac matrices (with $\mu = 0$ the temporal component and $\mu = 1, 2$ spatial components), the ellipsis denotes additional allowed contributions such as the Maxwell term $\sim (\epsilon_{\mu\nu\kappa} \partial_\nu a_\kappa)^2$, and $A_\mu$ is the electromagnetic vector potential. The two-component spinor $\Psi$ thus carries electric charge while $\tilde \Psi$ is neutral; in the QED$_3$ formulation electric currents are instead encoded through fluxes of the dynamical gauge field $a_\mu$. 

Equation~\eqref{eq.freedirac} captures two setups of interest here. $(i)$~Without an applied magnetic field ($B \equiv \partial_1 A_2 - \partial_2 A_1 = 0$), the theory describes the time-reversal-preserving Dirac cone routinely observed in TI surfaces. In this context Eq.~\eqref{eqn.qed3} provides an equivalent surface description in which the emergence of certain strongly interacting symmetric gapped phases \cite{BondersonTO,ChenTO,MetlitskiTO,WangTO} becomes extremely natural \cite{WangSenthil2015,MetlitskiVishwanath2015}. $(ii)$~Alternatively, one can view Eq.~\eqref{eq.freedirac} as a Dirac theory enjoying an exact particle-hole symmetry \cite{MetlitskiVishwanath2015,WangSenthilReview} that pins the chemical potential to the Dirac point ($A_0 = 0$) while permitting a finite magnetic field ($B \neq 0$) \footnote{Such a particle-hole symmetric Dirac theory can arise at the surface of a class AIII 3D topological superconductor \cite{MetlitskiVishwanath2015,WangSenthilReview}.}. The field rearranges the spectrum into Landau levels symmetric about zero energy. In particular, one level sits exactly at zero energy and is constrained to half-filling by particle-hole symmetry. At low energies the system maps precisely onto the half-filled lowest Landau level in a conventional 2D electron gas; interactions can thus generate a CFL. For this high-field problem Eq.~\eqref{eqn.qed3} describes the particle-hole-symmetric CFL reformulation introduced recently by Son \cite{Son}, with $\tilde \Psi$ interpreted as the appropriate composite fermion field; see also \cite{WangSenthilReview,Geraedts,ShankarMurthy,Kachru}.

Two methods have been employed to support the duality between Eqs.~\eqref{eq.freedirac} and \eqref{eqn.qed3} \cite{WangSenthil2015,MetlitskiVishwanath2015,WangSenthilReview,metlitskiduality}. First, various known TI surface phases were shown to be accessible in either formulation; second, an electric-magnetic duality for a (gauged) TI \emph{bulk} was shown to recover the alternate surface theory in Eq.~\eqref{eqn.qed3}. We develop a new operator-based derivation of this duality \emph{directly in two spatial dimensions} by relating the fermions $\Psi$ and $\tilde{\Psi}$ and explicitly mapping the path integrals onto one another. As we will see, the emergent gauge field $a_\mu$ reflects the nonlocal relation between the two fermion fields. %Our essentially exact mapping enables consistency checks for the duality and also leads to nontrivial predictions for operator scaling dimensions in QED$_3$. 
Our essentially exact mapping allows us to demonstrate that QED$_3$ regularized as in our construction shares the same low-energy behavior as a free Dirac cone, leading to nontrivial predictions for operator scaling dimensions in QED$_3$. We bolster this conclusion with several consistency checks (see Appendix). Furthermore, the mapping provides a derivation of Son's proposed CFL field theory \cite{Son}. That is, the half-filled Landau level problem that occurs in ${\cal S}_{\text{Dirac}}$ for $A_0 = 0$, $B \neq 0$ maps to dual fermions $\tilde \Psi$ coupled to a dynamical gauge field without a Chern-Simons term, and where the dual fermions are doped to a density matching the original magnetic flux. Generalizing the method to \emph{bosonic} TI surfaces \cite{VishwanathSenthil2013} allows us to obtain a dual description given by QED$_3$ with $N = 2$ fermion flavors. 

\begin{figure}
\includegraphics[width=\columnwidth]{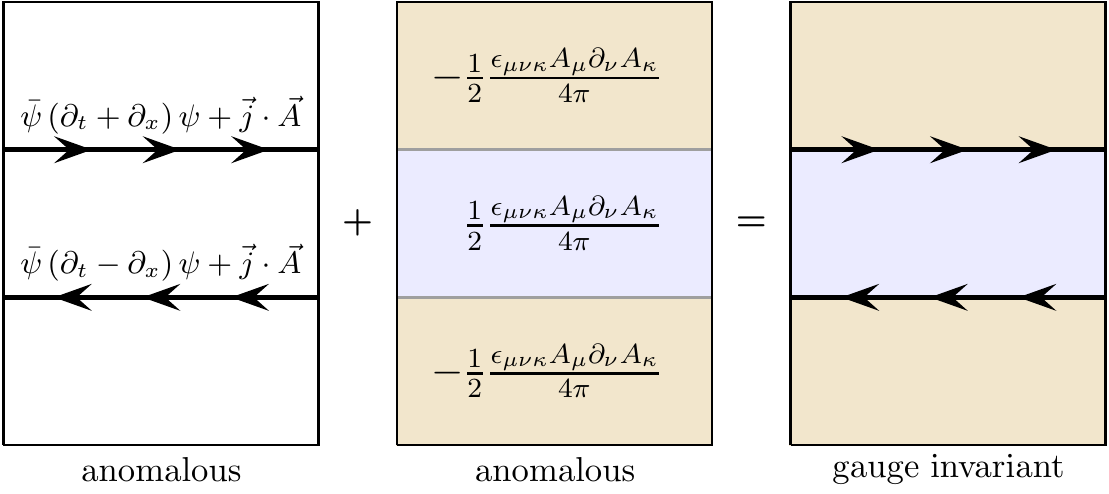}
\caption{An array of chiral wires (left) exhibits a gauge anomaly when coupled to the electromagnetic vector potential $A_\mu$. The same anomaly but with opposite sign arises in a 2D Chern-Simons theory with alternating coefficient (middle). The anomalies exactly cancel in the full 2D theory (right) containing both the chiral wires and the bulk Chern-Simons theory.}
\label{fig:anomaly}
\end{figure}

{\bf \emph{Model and symmetries.}}~Our analysis begins from an array of 1D chiral-electron `wires,' sketched in Fig.~\ref{fig:anomaly}, with Hamiltonian $H =\int_x \sum_y( h_\text{wire} + h_\text{hop})$ given by
\begin{align}
&h_\text{wire} = v(-1)^y \psi_y^\dagger(-i \partial_x)\psi_y ~, \label{Hintrawire} \\
&h_\text{hop} = -w (-1)^y
 (\psi_y^\dag \psi_{y+1} + \mathrm{H.c.}) ~. \label{Hhop}
\end{align}
Here $\psi^\dagger_y(x)$ creates a chiral electron at coordinate $x$ in wire $y$; electrons move rightward with speed $v$ for even $y$ and leftward for odd $y$. Equation~\eqref{Hintrawire} describes the intra-wire kinetic energy while Eq.~\eqref{Hhop} encodes inter-wire hopping with strength $w$. (Coupling to the vector potential $A_\mu$ will be discussed shortly.) The Hamiltonian manifestly conserves charge and is invariant under the \emph{antiunitary} symmetries ${\cal T}, {\cal C}$ defined as
\begin{equation}
{\cal T} \psi_y {\cal T}^{-1} = (-1)^y\psi_{y+1} ~, \ \ \ \ \ \ \ \ 
{\cal C} \psi_y {\cal C}^{-1} = \psi^\dagger_{y+1} ~.\label{eq.latticec}
\end{equation}

The band structure derived from Eqs.~\eqref{Hintrawire} and \eqref{Hhop} supports a single Dirac cone centered about zero momentum. Upon defining a slowly varying fermionic spinor $\Psi(x, y) \equiv [\psi_{2y}(x), \psi_{2y+1}(x)]^T$ one can readily take the continuum limit, which yields an effective Dirac Hamiltonian $H_\text{eff} = \int_{x,y}\Psi^\dagger \left[v \sigma^z (-i\partial_x) + w \sigma^y (-i\partial_y) \right] \Psi$ corresponding to Eq.~\eqref{eq.freedirac} ($\sigma^{x,y,z}$ are Pauli matrices; the associated Dirac matrices are $\gamma^0 = \sigma^x, \gamma^1 = -i\sigma^y, \gamma^2 = i\sigma^z$). Moreover, symmetries act on the continuum fields according to
\begin{equation}
 {\cal T} \Psi {\cal T}^{-1} = i \sigma^y \Psi ~,\ \ \ \ \ \ \ \ \ \ \ \ \ 
 {\cal C} \Psi {\cal C}^{-1} = \sigma^x \Psi^\dagger ~.\label{eq.contc}
\end{equation}
Viewed from the low-energy limit, $\mathcal{T}$ reproduces the time-reversal transformation familiar from the TI surface while $\mathcal{C}$ is the particle-hole symmetry relevant for the half-filled Landau level setting. Our wire model therefore captures both systems of interest---but in a way that facilitates an explicit duality transformation.

When coupling the wire model Eqs.~\eqref{Hintrawire}-\eqref{Hhop} to the external electromagnetic vector potential $A_\mu$, some care must be taken to ensure gauge invariance. Each chiral electron wire is on its own anomalous and cannot be made gauge invariant in a purely $(1+1)$-dimensional system. This situation is familiar from the chiral edge state of a quantum Hall system. There, gauge invariance is maintained due to the bulk Chern-Simons term for the vector potential~\cite{XGanomaly}. For the present case, an alternating Chern-Simons term for $A_\mu$ between wires,
\begin{align}
 {\cal S}_\text{staggered-CS} = -\sum_y\frac{1}{2}\frac{(-1)^y}{4\pi} \int_{t,x}\int_{y}^{y+1}d y' \epsilon_{\mu\nu\kappa}A_\mu \partial_\nu A_\kappa ~,
 \label{eq.lprime}
\end{align}
ensures gauge invariance as Fig.~\ref{fig:anomaly} sketches. 
See Appendix for details in the wire model.
The fractional Chern-Simons coefficient can be understood, e.g., if one views the wire model as describing a $\mathcal{T}$-symmetric `antiferromagnetic' TI \cite{AFTImong,AFTIessin,AFTIfang} surface decorated with alternating magnetic strips that generate a staggered Hall conductance $\sigma_{xy} = \pm e^2/(2h)$. In the continuum limit this oscillating term drops out and we obtain ${\cal S}_\text{Dirac}$ in Eq.~\eqref{eq.freedirac}.

{\bf \emph{Dual fermions}.}~To perform the duality we adopt a bosonized description, writing 
\begin{equation}
\psi_y(x) = \eta_y e^{i \phi_y(x)}~,~~~~~~~~(\text{electron})
\label{psiy}
\end{equation}
where $\phi_y(x)$ are chiral boson fields satisfying
$[\phi_y(x), \phi_{y'}(x')] = \delta_{yy'} (-1)^y \,i\, \pi\, \text{sgn}(x-x')$ 
and $\eta_y$ are Klein factors ensuring anticommutation between fermions in different wires. We define \emph{dual} fields via
\begin{equation}
\tilde{\phi}_y \equiv \sum_{y'\neq y} \text{sgn}\(y-y'\) (-1)^{y'}\phi_{y'}
\equiv \sum_{y'} D_{y,y'}\phi_{y'} ~.
\label{tildephi}
\end{equation}
On the right side we introduced matrix notation $D_{y, y'} = (1 - \delta_{yy'}) \text{sgn}\left(y-y'\right) (-1)^{y'}$ for later convenience.
The dual fields obey
\begin{align}
[ \tilde{\phi}_y(x), \tilde{\phi}_{y'}(x')] = -[ \phi_y(x), \phi_{y'}(x')] ~,\label{eqn:dualdef}
\end{align}
i.e., they are also chiral bosons with chirality opposite to $\phi_y$. We can therefore define a new \emph{dual fermion} operator as
\begin{align}
\tilde \psi_y = \eta_y (-1)^{y(y+1)/2} e^{i \tilde{\phi}_y}~,~~~~~~~~(\text{dual fermion})
\end{align}
using the same Klein factors from Eq.~\eqref{psiy}. (The $y$-dependent phase inserted above merely simplifies expressions that follow.) 
 This dual fermion is a key result of this paper and exhibits the following properties:

\begin{enumerate}

\item Applying duality to $\tilde{\psi}_y$ recovers the original electrons [i.e., $D^2 = 1$ for the matrix defined in Eq.~(\ref{tildephi})].

\item Upon encircling an electron, the dual fermion acquires a phase of $4\pi$ (see below).

 \item Dual fermion hopping between wires is a local process since
 \begin{align}
 \tilde \psi_y^\dagger \tilde\psi_{y+1} = \begin{cases}
 \psi_{y+1}^\dagger\psi_y ~,& \text{for}~y~\text {even}\\
 \psi_y^\dagger \psi_{y+1} ~,& \text{for}~y~\text {odd}.
\label{dualhop}
\end{cases}
\end{align}
Hence the original electron hopping Hamiltonian in Eq.~\eqref{Hhop} maps to a dual fermion hopping Hamiltonian of identical form.
Together with the chiral nature of $\tilde{\psi}_y$,
this motivates defining slowly varying \emph{dual Dirac fermions} $\tilde{\Psi}(x,y)=[\tilde{\psi}_{2y}(x), \tilde{\psi}_{2y+1}(x)]^T$ analogous to $\Psi(x,y)$. 

\item Time reversal and charge conjugation act locally on the dual Dirac fermions, but their roles are interchanged compared to electrons:
\begin{equation}
{\cal T} \tilde{\Psi} {\cal T}^{-1} = \sigma^x \tilde{\Psi}^\dagger ~,\ \ \ \ \ \ \ \ \ \
{\cal C} \tilde{\Psi} {\cal C}^{-1} = i \sigma^y \tilde{\Psi} ~, \label{dualsymmetries}
\end{equation}
(up to unimportant overall phase factors); cf.~Eq.~\eqref{eq.contc}. 

\item Density $\tilde{\rho}_y = \frac{-(-1)^y}{2\pi} \partial_x \tilde{\phi}_y $ and intra-wire kinetic energy $\propto (-1)^y \tilde{\psi}_y^\dagger i \partial_x \tilde \psi_y \sim (\partial_x \tilde{\phi}_y)^2$ of dual fermions are non-local in terms of electrons and vice versa. This is the origin of an emergent gauge field $a_\mu$ in the dual formulation.
\end{enumerate}

\begin{figure}[ht]
\includegraphics[width=\columnwidth]{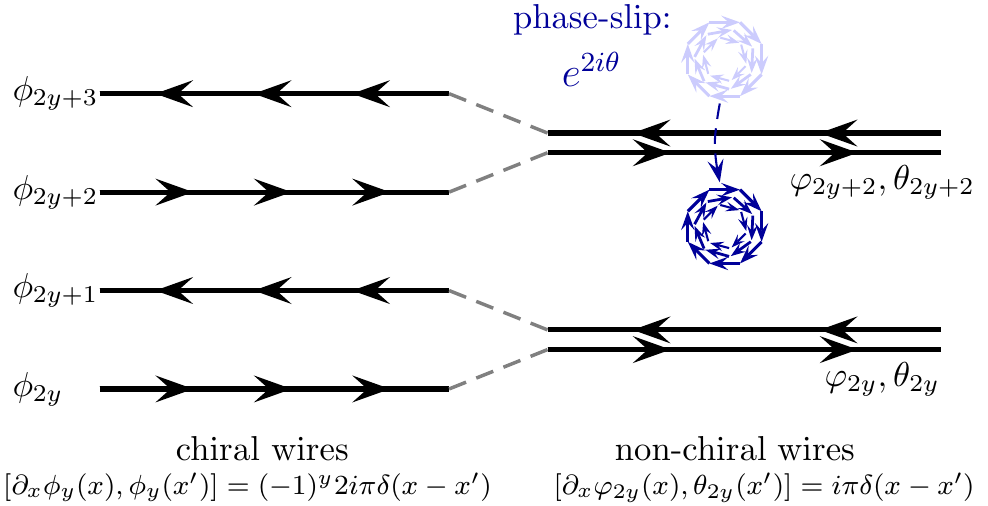}
\caption{Array of wires with alternating chirality (left) viewed as an array of non-chiral wires (right). In the non-chiral basis $\psi_{2y+1}^\dagger \psi_{2y} \sim e^{2 i \theta_{2y}}$ is the familiar phase-slip operator that one can view as tunneling a vortex.
} 
\label{fig:slip}
\end{figure}

To physically interpret the dual fermion, imagine combining pairs of counterpropagating chiral wires into non-chiral ones as in Fig.~\ref{fig:slip}. For each non-chiral wire we define canonical boson fields $\varphi_{2y},\theta_{2y} = (\phi_{2y} \pm \phi_{2y+1})/2$, where $e^{2 i \theta_{2y}}$ introduces a $2\pi$ phase slip in $\varphi_{2y}$. Viewing each non-chiral wire as a (power-law) superconductor, $e^{2i \theta_{2y}}$ equivalently produces a $4\pi$ phase slip in the phase of the Cooper pair $e^{2i \varphi_{2y}}$ and thus hops an $hc/e$ vortex across the wire; see Fig.~\ref{fig:slip}. 
Dual fermions are expressed in this language as
\begin{equation}
 \tilde{\psi}^\dagger_{2y/2y+1} \sim \psi^\dagger_{2y+1/2y} \prod_{y' > y} e^{2 i \theta_{2y'}} \prod_{y' < y} e^{-2 i \theta_{2y'}} ~.
\end{equation}
The two products each describe an $hc/e$ vortex of the same sign moving in from (opposite) infinity. Thus, the dual fermion may be viewed as two $hc/e$ vortices tied to an electron, as in the very different approaches from \cite{MetlitskiVishwanath2015,WangSenthil2015}. We emphasize that the electron/vortex binding follows from our judicious change of variables, which underlies the remarkable properties of the dual fields enumerated above. The local action of $\mathcal{T}$ and $\mathcal{C}$ on the dual fermions is particularly noteworthy since this property is difficult to realize with the standard flux-attachment procedure. 

{\bf \emph{Duality of the path integral.}}~In bosonized language, the (real-time) path integral for the original electrons, assuming $A_\mu = 0$ for now, reads ${\cal Z} = \int {\cal D}\phi e^{i {\cal S}}$ with
\begin{equation}
{\cal S} = \int_{t,x}\sum_y\left\{\frac{(-1)^y \partial_x \phi_y \partial_t \phi_y}{4\pi} - h_{\text{wire}}[\phi] - h_\text{hop}[\phi]\right\}.
\label{L}
\end{equation}
As noted above, the inter-wire hopping term is self-dual, i.e., $h_\text{hop}[\phi] = h_\text{hop}[\tilde\phi]$. It further follows from Eq.~\eqref{eqn:dualdef}, or equivalently by changing variables in the path integral, that the time-derivative term only acquires a minus sign when mapping $\phi \rightarrow \tilde \phi$. Intra-wire kinetic energy does however transform nontrivially under duality: 
\begin{align}
h_{\text{wire}}=\frac{v}{4\pi} (\partial_x \phi_y)^2 = \frac{v}{4\pi} \bigg(\partial_x \sum_{y'} D_{y,y'}\tilde\phi_{y'}\bigg)^2 ~.
\label{Hwire}
\end{align}
Indeed, the expression on the right is highly non-local---which is not surprising from the viewpoint of the more familiar boson-vortex dualities, where dual vortices exhibit long-range interactions \cite{DasguptaHalperin,MatthewDungHai}.

To obtain a local theory, we now formally introduce new integration variables $a_{0, y}(x,t)$ and $a_{1, y}(x,t)$ and add two complete squares to the action by replacing 
\begin{align}
 &h_{\text{wire}}\rightarrow h_{\text{wire}} + \frac{1}{4\pi} \bigg{\{} (v+u)\(\partial_x\tilde\phi_{y}-a_{1,y}\)^2\label{shift}\\
 &- v\(2\Delta^{-1,T}[(-1)^y\partial_x \tilde \phi_y] + \frac{\Delta [a_{0,y}-v(-1)^y a_{1,y}]}{2v}\)^2 \bigg{\}}\nonumber
\end{align}
in Eq.~\eqref{L}. Here $\Delta$ denotes a lattice derivative, e.g., $\Delta a_{\mu, y} = a_{\mu, y+1} - a_{\mu, y}$; $\Delta^{-1,T}$ is its inverse transpose, i.e., $\sum_y \Delta^{-1,T}[a_{\mu,y}]\Delta[a_{\mu,y}] = \sum_y a_{\mu,y}^2$ which defines $\Delta^{-1,T}$ up to a constant. We take $\Delta^{-1,T}a_{\mu,y} = \frac{1}{2}\sum_{y'} {\rm sgn}(y'-y + 0^+)a_{\mu,y'}$. The shift \eqref{shift} multiplies the path integral by a benign constant but is nevertheless very useful. Given the expressions for $\Delta^{-1,T}$ and $D$, one readily verifies that all quadratic terms in $\partial_x \tilde \phi_y$ with coefficient $v$ exactly cancel---leaving only local contributions!

The remaining terms can be reorganized to obtain ${\cal Z}\sim \int {\cal D}\tilde \phi {\cal D}a_\mu e^{i {\cal S}_\text{dual}}$ with dual action
\begin{eqnarray}
{\cal S}_\text{dual} &=& \int_{t,x}\sum_y\bigg\{\frac{-(-1)^y \partial_x \tilde{\phi}_y \partial_t \tilde{\phi}_y}{4\pi} - h_{\text{wire}}^{\text{dual}}[\tilde{\phi},a_\mu] 
\nonumber \\
 &-& h_{\text{hop}}[\tilde{\phi}] \bigg\} + {\cal S}_\text{MW} [a_\mu]+ {\cal S}_\text{staggered-CS}'[a_\mu].\label{eqn.sdual} 
\end{eqnarray}
Intra-wire dual-fermion kinetic energy is described by
\begin{equation}
 h_{\text{wire}}^{\text{dual}}[\tilde \phi,a_\mu] = \frac{-(-1)^y}{2\pi} a_{0,y} \partial_x \tilde{\phi}_y +\frac{u}{4\pi}(\partial_x \tilde \phi_y -a_{1,y})^2 \nonumber,
\end{equation}
which has a similar form to Eq.~\eqref{Hwire} except that here the fermions minimally couple to the emergent gauge field $a_\mu$. A Maxwell term for the gauge field appears in
\begin{equation}
 {\cal S}_\text{MW}[a_\mu] = \int_{t,x} \sum_y\left[\frac{1}{16 \pi v}(\Delta a_{0,y})^2 - \frac{v}{16\pi}(\Delta a_{1,y})^2 \right]. \nonumber
\end{equation}
In the continuum limit we indeed recover an anisotropic Maxwell term ${\cal S}_\text{MW}\sim \lambda_\mu\left( \epsilon_{\mu\nu\kappa}\partial_\nu a_{\kappa}\right)^2$ in the $a_2 = 0$ gauge with $\lambda_y = 0$. Note that $\lambda_y=0$ is a property of the bare microscopic theory; finite $\lambda_y$ will be generated under renormalization with any non-zero inter-wire hopping amplitude $w$. Finally, ${\cal S}_\text{staggered-CS}'= \int_{t,x}\sum_y\frac{(-1)^y}{8\pi}\Delta a_{0,y} \left(a_{1,y+1} + a_{1,y}\right)$ is the discrete analogue of the staggered Chern-Simons coupling in Eq.~\eqref{eq.lprime}, but for $a_\mu$. This term similarly drops out in the continuum limit. 

Refermionizing ${\cal S}_\text{dual}$ and taking the continuum limit yields ${\cal S}_\text{{QED}$_3$}$ of Eq.~\eqref{eqn.qed3} in the special case $A_\mu = 0$.
In the Appendix, we perform the same analysis including the electromagnetic vector potential $A_\mu$. We find that in the continuum limit the dual action above merely acquires an extra mutual Chern-Simons term ${\cal S}_\text{CS}[a_\mu,A_\mu] = \int_{t,x,y}\frac{1}{4\pi}\epsilon_{\mu\nu\kappa}A_\mu \partial_\nu a_\kappa$, precisely as in Eq.~\eqref{eqn.qed3}. 
This completes our duality derivation---which we emphasize makes no assumption about the value of $A_\mu$. In particular, by virtue of the mutual Chern-Simons term, uniform chemical potential for electrons $A_0 \neq 0$ translates to a non-zero orbital magnetization for dual fermions, while electrons in magnetic field $B$ correspond to dual fermions at finite density $n_\text{dual}= \frac{1}{4\pi}B$. Our analysis thus immediately applies both to the TI surface \emph{and} half-filled Landau level.

{\bf \emph{Duality for gapped phases.}} As a first application we briefly discuss interaction-induced gapped phases descending from the Dirac cone/QED$_3$ theories at $B = 0$. References \onlinecite{WangSenthil2015,MetlitskiVishwanath2015} argued phenomenologically that a Fu-Kane superconductor \cite{FuKane} of dual fermions corresponds to `T-Pfaffian' non-Abelian topological order \cite{BondersonTO,ChenTO} for electrons and vice versa. Our explicit duality mapping allows us to readily translate the dual-fermion interactions needed for the former (which are straightforward to obtain) into an electronic Hamiltonian for the T-Pfaffian (which is highly nontrivial). 

We simply sketch the construction here; for details see the Appendix. First, within our wire formulation it proves convenient to enlarge the unit cell and relax the symmetries in Eq.~\eqref{eq.latticec} by enforcing only $\mathcal{T}^3$ and $\mathcal{C}^3$. One can then construct a coupled-wire Hamiltonian with three (electronic and dual) Dirac cones---two of which can be gapped without breaking symmetries. At low energies one thus again obtains the continuum theories \eqref{eq.freedirac} and \eqref{eqn.qed3} with symmetries acting according to Eqs.~\eqref{eq.contc} and \eqref{dualsymmetries}, but now with auxiliary gapped Dirac cones. Suppose that we specifically add interactions that gap the auxiliary \emph{dual} Dirac cones through a pairing instability that spontaneously breaks dual-fermion number conservation. This condensate induces a proximity effect on the remaining massless dual Dirac cone and drives the dual fermions into a Fu-Kane superconductor. Quite remarkably, dualizing these interactions yields precisely the electron Hamiltonian found in Ref.~\onlinecite{STO} to generate the T-Pfaffian for the electron TI surface. This provides a strong consistency check on our analysis.

{\bf \emph{Properties of $N=1$ QED$_3$.}}
Our explicit duality mapping also allows nontrivial predictions for QED$_3$, a naively strongly interacting theory, by relating operators to their free-electron counterparts. As an important example, 
Eq.~(\ref{dualhop}) immediately gives
\begin{equation}
 m \int_x\sum_y(\tilde{\psi}_y^\dagger \tilde{\psi}_{y+1} + \text{H.c.}) = m \int_x\sum_y(\psi_y^\dagger \psi_{y+1} + \text{H.c.}) ~.
\label{massterms}
\end{equation}
In the continuum limit 
the left and right sides respectively yield the dual-fermion mass term $m\tilde{\Psi}^\dagger \sigma^1 \tilde{\Psi}$ and electron mass term $m\Psi^\dagger \sigma^1 \Psi$, and hence the two are identified! It follows that the scaling dimension for the mass in QED$_3$ must be \emph{precisely} 2. To appreciate this result, note that in QED$_3$ with $N$ fermion flavors, a large-$N$ treatment predicts a positive $O(1/N)$ correction to the free-fermion result arising from gauge-field-mediated interactions \cite{Hermele_ASL_Erratum_2007}. 
Our results imply that such corrections must exactly cancel when summed to all orders in $1/N$ in the $N = 1$ limit---even though the fermions in QED$_3$ strongly couple to the gauge field. In the Appendix we additionally explore analogues of Eq.~\eqref{massterms} for fermion currents. Constraints from current conservation allow us to make exact statements about scaling dimensions in this case, providing further consistency checks on the duality.

{\bf \emph{Duality for $N=2$ QED$_3$.}}~ 
It is interesting to ask whether similar mappings exist for larger-$N$ QED$_3$. To address this question, and illustrate our method's generality, we run our formalism `in reverse' to determine the theory dual to $N=2$ QED$_3$. The latter corresponds to two copies of the wire model \eqref{eqn.sdual} labelled by $\sigma = \pm$, each with its own species of dual fermions $\tilde \psi_\sigma \sim e^{i \tilde \phi_\sigma}$ but with the \emph{same} dynamical gauge field $a_\mu$.
The intra-wire Hamiltonian density expressed in terms of (co-propagating) charge and neutral 
modes $\tilde \phi_{c/n} =( \tilde \phi_+ \pm \tilde \phi_-)/2$ is $h^\text{dual}_\text{wire}[\tilde \phi_c,a_\mu] + h_\text{wire}[\tilde \phi_n]$. (Here `charge' and `neutral' are with respect to the dynamical gauge field $a_\mu$ rather than the external vector potential $A_\mu$.) 
Since only $\tilde \phi_c$ couples to $a_\mu$ we implement the duality by defining a chiral dual charge mode $\phi_{c,y} =\sum_{y'} D_{y,y'}\tilde\phi_{c,y'}$ that counterpropagates relative to the neutral mode $\tilde \phi_{n,y}$---which we leave intact. Integrating out $a_\mu$ then yields a \emph{local} theory for $\phi_c$ with intra-wire kinetic energy $h_\text{wire}[\phi_c]$. Inter-wire hopping becomes [cf.~Eq.~\eqref{dualhop}]
\begin{equation}
\sum\nolimits_\sigma \tilde\psi_{\sigma,y}^\dagger \tilde\psi_{\sigma,y+1}+\mathrm{H.c.}\rightarrow \sum\nolimits_\sigma b_{\sigma,y}^\dagger b_{\sigma,y+1}+\mathrm{H.c.}\label{eqn.bosonhop},
\end{equation}
where $b_{\sigma,y}^\dagger\sim e^{-i\phi_{c,y} -\sigma i \tilde\phi_{n,y}}$ creates a \emph{dual boson} of species $\sigma$. This suggests introducing non-chiral modes $\varphi_{\sigma,y}= (\phi_{c,y} +\sigma \tilde\phi_{n,y})$ which obey the intra-wire action
\small
\begin{align}
& {\cal S}_\text{wire} = \int_{t,x} \sum_{y}\left[\frac{(-1)^yK_{\sigma\sigma'}\partial_x \varphi_{\sigma,y}\partial_t \varphi_{\sigma',y}}{4\pi}+\frac{v_{\sigma}}{4 \pi}\(\partial_x \varphi_{\sigma,y}\)^2\right]\label{eqn.bosonwire}
\end{align}
\normalsize
with K-matrix $K=\sigma^x$. 

The network model described by Eqs.~\eqref{eqn.bosonhop} and \eqref{eqn.bosonwire} was introduced in Ref.~\onlinecite{VishwanathSenthil2013} to describe the surface of time-reversal-invariant bosonic TIs with two separately conserved U(1) symmetries. We now see explicitly that $N = 2$ QED$_3$ provides a dual description of the bosonic TI surface \footnote{In fact, $N=2$ QED$_3$ exhibits additional symmetries whose action on the dual bosons $b_{\sigma,y}$ is derived in the Appendix.}. This duality was proposed based on bulk arguments in Refs.~\onlinecite{wangsenthilu1,XuYouSPT}. Unlike the duality for $N=1$ QED$_3$, however, the $N=2$ analogue does not map onto a simple free theory. Thus one cannot readily infer operator scaling dimensions in this gauge theory by mapping between the two. Inspired by its utility in the fermion case, we nevertheless expect that the duality will be conceptually useful, e.g., for accessing novel surface phases of bosonic TIs.

{\bf \emph{Conclusions.}}
We derived the equivalence of a free Dirac cone and $N=1$ QED$_{3}$ on the level of the path integral. We used this mapping to determine the scaling dimension of the fermion mass in $N=1$ QED$_{3}$, and illustrated with the example of the T-Pfaffian how this formulation can be used to easily obtain electron Hamiltonians for topologically ordered phases. By running the same mapping `in reverse' we found that $N=2$ QED$_{3}$ is dual to a critical theory relevant for bosonic TI surfaces. We expect that generalizations of our approach will provide insights into other exotic phases with emergent gauge fields, such as gapless quantum spin liquids.

{\bf \emph{Acknowledgments.}} We gratefully acknowledge M.~Mulligan, S.~Raghu, T.~Senthil, A.~Vishwanath and C.~Xu for valuable discussions.
This work was supported by the NSF through grant DMR-1341822 (JA) and grant DMR-1206096 (OIM); the Alfred P. Sloan Foundation (JA); the Caltech Institute for Quantum Information and Matter, an NSF Physics Frontiers Center with support of the Gordon and Betty Moore Foundation; and the Walter Burke Institute for Theoretical Physics at Caltech.

\bibliography{TIdual}

%merlin.mbs apsrev4-1.bst 2010-07-25 4.21a (PWD, AO, DPC) hacked
%Control: key (0)
%Control: author (8) initials jnrlst
%Control: editor formatted (1) identically to author
%Control: production of article title (-1) disabled
%Control: page (0) single
%Control: year (1) truncated
%Control: production of eprint (0) enabled
\begin{thebibliography}{44}%
\makeatletter
\providecommand \@ifxundefined [1]{%
 \@ifx{#1\undefined}
}%
\providecommand \@ifnum [1]{%
 \ifnum #1\expandafter \@firstoftwo
 \else \expandafter \@secondoftwo
 \fi
}%
\providecommand \@ifx [1]{%
 \ifx #1\expandafter \@firstoftwo
 \else \expandafter \@secondoftwo
 \fi
}%
\providecommand \natexlab [1]{#1}%
\providecommand \enquote  [1]{``#1''}%
\providecommand \bibnamefont  [1]{#1}%
\providecommand \bibfnamefont [1]{#1}%
\providecommand \citenamefont [1]{#1}%
\providecommand \href@noop [0]{\@secondoftwo}%
\providecommand \href [0]{\begingroup \@sanitize@url \@href}%
\providecommand \@href[1]{\@@startlink{#1}\@@href}%
\providecommand \@@href[1]{\endgroup#1\@@endlink}%
\providecommand \@sanitize@url [0]{\catcode `\\12\catcode `\$12\catcode
  `\&12\catcode `\#12\catcode `\^12\catcode `\_12\catcode `\%12\relax}%
\providecommand \@@startlink[1]{}%
\providecommand \@@endlink[0]{}%
\providecommand \url  [0]{\begingroup\@sanitize@url \@url }%
\providecommand \@url [1]{\endgroup\@href {#1}{\urlprefix }}%
\providecommand \urlprefix  [0]{URL }%
\providecommand \Eprint [0]{\href }%
\providecommand \doibase [0]{http://dx.doi.org/}%
\providecommand \selectlanguage [0]{\@gobble}%
\providecommand \bibinfo  [0]{\@secondoftwo}%
\providecommand \bibfield  [0]{\@secondoftwo}%
\providecommand \translation [1]{[#1]}%
\providecommand \BibitemOpen [0]{}%
\providecommand \bibitemStop [0]{}%
\providecommand \bibitemNoStop [0]{.\EOS\space}%
\providecommand \EOS [0]{\spacefactor3000\relax}%
\providecommand \BibitemShut  [1]{\csname bibitem#1\endcsname}%
\let\auto@bib@innerbib\@empty
%</preamble>
\bibitem [{\citenamefont {Fu}\ \emph {et~al.}(2007)\citenamefont {Fu},
  \citenamefont {Kane},\ and\ \citenamefont {Mele}}]{FuTI}%
  \BibitemOpen
  \bibfield  {author} {\bibinfo {author} {\bibfnamefont {L.}~\bibnamefont
  {Fu}}, \bibinfo {author} {\bibfnamefont {C.~L.}\ \bibnamefont {Kane}}, \ and\
  \bibinfo {author} {\bibfnamefont {E.~J.}\ \bibnamefont {Mele}},\ }\href
  {\doibase 10.1103/PhysRevLett.98.106803} {\bibfield  {journal} {\bibinfo
  {journal} {Phys. Rev. Lett.}\ }\textbf {\bibinfo {volume} {98}},\ \bibinfo
  {pages} {106803} (\bibinfo {year} {2007})}\BibitemShut {NoStop}%
\bibitem [{\citenamefont {Moore}\ and\ \citenamefont
  {Balents}(2007)}]{MooreTI}%
  \BibitemOpen
  \bibfield  {author} {\bibinfo {author} {\bibfnamefont {J.~E.}\ \bibnamefont
  {Moore}}\ and\ \bibinfo {author} {\bibfnamefont {L.}~\bibnamefont
  {Balents}},\ }\href {\doibase 10.1103/PhysRevB.75.121306} {\bibfield
  {journal} {\bibinfo  {journal} {Phys. Rev. B}\ }\textbf {\bibinfo {volume}
  {75}},\ \bibinfo {pages} {121306} (\bibinfo {year} {2007})}\BibitemShut
  {NoStop}%
\bibitem [{\citenamefont {Roy}(2009)}]{RoyTI}%
  \BibitemOpen
  \bibfield  {author} {\bibinfo {author} {\bibfnamefont {R.}~\bibnamefont
  {Roy}},\ }\href {\doibase 10.1103/PhysRevB.79.195322} {\bibfield  {journal}
  {\bibinfo  {journal} {Phys. Rev. B}\ }\textbf {\bibinfo {volume} {79}},\
  \bibinfo {pages} {195322} (\bibinfo {year} {2009})}\BibitemShut {NoStop}%
\bibitem [{\citenamefont {Hasan}\ and\ \citenamefont
  {Kane}(2010)}]{KaneReview}%
  \BibitemOpen
  \bibfield  {author} {\bibinfo {author} {\bibfnamefont {M.~Z.}\ \bibnamefont
  {Hasan}}\ and\ \bibinfo {author} {\bibfnamefont {C.~L.}\ \bibnamefont
  {Kane}},\ }\href {\doibase 10.1103/RevModPhys.82.3045} {\bibfield  {journal}
  {\bibinfo  {journal} {Rev. Mod. Phys.}\ }\textbf {\bibinfo {volume} {82}},\
  \bibinfo {pages} {3045} (\bibinfo {year} {2010})}\BibitemShut {NoStop}%
\bibitem [{\citenamefont {Qi}\ and\ \citenamefont {Zhang}(2011)}]{QiReview}%
  \BibitemOpen
  \bibfield  {author} {\bibinfo {author} {\bibfnamefont {X.-L.}\ \bibnamefont
  {Qi}}\ and\ \bibinfo {author} {\bibfnamefont {S.-C.}\ \bibnamefont {Zhang}},\
  }\href {\doibase 10.1103/RevModPhys.83.1057} {\bibfield  {journal} {\bibinfo
  {journal} {Rev. Mod. Phys.}\ }\textbf {\bibinfo {volume} {83}},\ \bibinfo
  {pages} {1057} (\bibinfo {year} {2011})}\BibitemShut {NoStop}%
\bibitem [{\citenamefont {Halperin}\ \emph {et~al.}(1993)\citenamefont
  {Halperin}, \citenamefont {Lee},\ and\ \citenamefont {Read}}]{HLR}%
  \BibitemOpen
  \bibfield  {author} {\bibinfo {author} {\bibfnamefont {B.~I.}\ \bibnamefont
  {Halperin}}, \bibinfo {author} {\bibfnamefont {P.~A.}\ \bibnamefont {Lee}}, \
  and\ \bibinfo {author} {\bibfnamefont {N.}~\bibnamefont {Read}},\ }\href
  {\doibase 10.1103/PhysRevB.47.7312} {\bibfield  {journal} {\bibinfo
  {journal} {Phys. Rev. B}\ }\textbf {\bibinfo {volume} {47}},\ \bibinfo
  {pages} {7312} (\bibinfo {year} {1993})}\BibitemShut {NoStop}%
\bibitem [{\citenamefont {Willett}\ \emph {et~al.}(1990)\citenamefont
  {Willett}, \citenamefont {Paalanen}, \citenamefont {Ruel}, \citenamefont
  {West}, \citenamefont {Pfeiffer},\ and\ \citenamefont {Bishop}}]{WillettCFL}%
  \BibitemOpen
  \bibfield  {author} {\bibinfo {author} {\bibfnamefont {R.~L.}\ \bibnamefont
  {Willett}}, \bibinfo {author} {\bibfnamefont {M.~A.}\ \bibnamefont
  {Paalanen}}, \bibinfo {author} {\bibfnamefont {R.~R.}\ \bibnamefont {Ruel}},
  \bibinfo {author} {\bibfnamefont {K.~W.}\ \bibnamefont {West}}, \bibinfo
  {author} {\bibfnamefont {L.~N.}\ \bibnamefont {Pfeiffer}}, \ and\ \bibinfo
  {author} {\bibfnamefont {D.~J.}\ \bibnamefont {Bishop}},\ }\href {\doibase
  10.1103/PhysRevLett.65.112} {\bibfield  {journal} {\bibinfo  {journal} {Phys.
  Rev. Lett.}\ }\textbf {\bibinfo {volume} {65}},\ \bibinfo {pages} {112}
  (\bibinfo {year} {1990})}\BibitemShut {NoStop}%
\bibitem [{\citenamefont {Kang}\ \emph {et~al.}(1993)\citenamefont {Kang},
  \citenamefont {Stormer}, \citenamefont {Pfeiffer}, \citenamefont {Baldwin},\
  and\ \citenamefont {West}}]{KangCFL}%
  \BibitemOpen
  \bibfield  {author} {\bibinfo {author} {\bibfnamefont {W.}~\bibnamefont
  {Kang}}, \bibinfo {author} {\bibfnamefont {H.~L.}\ \bibnamefont {Stormer}},
  \bibinfo {author} {\bibfnamefont {L.~N.}\ \bibnamefont {Pfeiffer}}, \bibinfo
  {author} {\bibfnamefont {K.~W.}\ \bibnamefont {Baldwin}}, \ and\ \bibinfo
  {author} {\bibfnamefont {K.~W.}\ \bibnamefont {West}},\ }\href {\doibase
  10.1103/PhysRevLett.71.3850} {\bibfield  {journal} {\bibinfo  {journal}
  {Phys. Rev. Lett.}\ }\textbf {\bibinfo {volume} {71}},\ \bibinfo {pages}
  {3850} (\bibinfo {year} {1993})}\BibitemShut {NoStop}%
\bibitem [{\citenamefont {Goldman}\ \emph {et~al.}(1994)\citenamefont
  {Goldman}, \citenamefont {Su},\ and\ \citenamefont {Jain}}]{GoldmanCFL}%
  \BibitemOpen
  \bibfield  {author} {\bibinfo {author} {\bibfnamefont {V.~J.}\ \bibnamefont
  {Goldman}}, \bibinfo {author} {\bibfnamefont {B.}~\bibnamefont {Su}}, \ and\
  \bibinfo {author} {\bibfnamefont {J.~K.}\ \bibnamefont {Jain}},\ }\href
  {\doibase 10.1103/PhysRevLett.72.2065} {\bibfield  {journal} {\bibinfo
  {journal} {Phys. Rev. Lett.}\ }\textbf {\bibinfo {volume} {72}},\ \bibinfo
  {pages} {2065} (\bibinfo {year} {1994})}\BibitemShut {NoStop}%
\bibitem [{\citenamefont {Sarma}\ and\ \citenamefont
  {Pinczuk}(1996)}]{DasSarmaBook}%
  \BibitemOpen
  \bibfield  {author} {\bibinfo {author} {\bibfnamefont {S.~D.}\ \bibnamefont
  {Sarma}}\ and\ \bibinfo {author} {\bibfnamefont {A.}~\bibnamefont
  {Pinczuk}},\ }\href@noop {} {\emph {\bibinfo {title} {{Perspectives in
  quantum Hall effects : novel quantum liquids in low-dimensional semiconductor
  structures}}}}\ (\bibinfo  {publisher} {Wiley},\ \bibinfo {address}
  {Chichester},\ \bibinfo {year} {1996})\BibitemShut {NoStop}%
\bibitem [{\citenamefont {Willett}(1997)}]{Willet}%
  \BibitemOpen
  \bibfield  {author} {\bibinfo {author} {\bibfnamefont {R.~L.}\ \bibnamefont
  {Willett}},\ }\href {\doibase 10.1080/00018739700101528} {\bibfield
  {journal} {\bibinfo  {journal} {Advances in Physics}\ }\textbf {\bibinfo
  {volume} {46}},\ \bibinfo {pages} {447} (\bibinfo {year} {1997})},\ \Eprint
  {http://arxiv.org/abs/http://dx.doi.org/10.1080/00018739700101528}
  {http://dx.doi.org/10.1080/00018739700101528} \BibitemShut {NoStop}%
\bibitem [{\citenamefont {Jain}(2007)}]{JainBook}%
  \BibitemOpen
  \bibfield  {author} {\bibinfo {author} {\bibfnamefont {J.~K.}\ \bibnamefont
  {Jain}},\ }\href@noop {} {\emph {\bibinfo {title} {{Composite Fermions}}}}\
  (\bibinfo  {publisher} {Cambridge University Press},\ \bibinfo {address}
  {Cambridge},\ \bibinfo {year} {2007})\BibitemShut {NoStop}%
\bibitem [{\citenamefont {Bonderson}\ \emph {et~al.}(2013)\citenamefont
  {Bonderson}, \citenamefont {Nayak},\ and\ \citenamefont {Qi}}]{BondersonTO}%
  \BibitemOpen
  \bibfield  {author} {\bibinfo {author} {\bibfnamefont {P.}~\bibnamefont
  {Bonderson}}, \bibinfo {author} {\bibfnamefont {C.}~\bibnamefont {Nayak}}, \
  and\ \bibinfo {author} {\bibfnamefont {X.-L.}\ \bibnamefont {Qi}},\ }\href
  {\doibase 10.1088/1742-5468/2013/09/P09016} {\bibfield  {journal} {\bibinfo
  {journal} {Journal of Statistical Mechanics: Theory and Experiment}\ }\textbf
  {\bibinfo {volume} {2013}},\ \bibinfo {pages} {P09016} (\bibinfo {year}
  {2013})}\BibitemShut {NoStop}%
\bibitem [{\citenamefont {Chen}\ \emph {et~al.}(2014)\citenamefont {Chen},
  \citenamefont {Fidkowski},\ and\ \citenamefont {Vishwanath}}]{ChenTO}%
  \BibitemOpen
  \bibfield  {author} {\bibinfo {author} {\bibfnamefont {X.}~\bibnamefont
  {Chen}}, \bibinfo {author} {\bibfnamefont {L.}~\bibnamefont {Fidkowski}}, \
  and\ \bibinfo {author} {\bibfnamefont {A.}~\bibnamefont {Vishwanath}},\
  }\href {\doibase 10.1103/PhysRevB.89.165132} {\bibfield  {journal} {\bibinfo
  {journal} {Phys. Rev. B}\ }\textbf {\bibinfo {volume} {89}},\ \bibinfo
  {pages} {165132} (\bibinfo {year} {2014})}\BibitemShut {NoStop}%
\bibitem [{\citenamefont {Metlitski}\ \emph {et~al.}(2015)\citenamefont
  {Metlitski}, \citenamefont {Kane},\ and\ \citenamefont
  {Fisher}}]{MetlitskiTO}%
  \BibitemOpen
  \bibfield  {author} {\bibinfo {author} {\bibfnamefont {M.~A.}\ \bibnamefont
  {Metlitski}}, \bibinfo {author} {\bibfnamefont {C.~L.}\ \bibnamefont {Kane}},
  \ and\ \bibinfo {author} {\bibfnamefont {M.~P.~A.}\ \bibnamefont {Fisher}},\
  }\href {\doibase 10.1103/PhysRevB.92.125111} {\bibfield  {journal} {\bibinfo
  {journal} {Phys. Rev. B}\ }\textbf {\bibinfo {volume} {92}},\ \bibinfo
  {pages} {125111} (\bibinfo {year} {2015})}\BibitemShut {NoStop}%
\bibitem [{\citenamefont {Wang}\ \emph {et~al.}(2013)\citenamefont {Wang},
  \citenamefont {Potter},\ and\ \citenamefont {Senthil}}]{WangTO}%
  \BibitemOpen
  \bibfield  {author} {\bibinfo {author} {\bibfnamefont {C.}~\bibnamefont
  {Wang}}, \bibinfo {author} {\bibfnamefont {A.~C.}\ \bibnamefont {Potter}}, \
  and\ \bibinfo {author} {\bibfnamefont {T.}~\bibnamefont {Senthil}},\ }\href
  {\doibase 10.1103/PhysRevB.88.115137} {\bibfield  {journal} {\bibinfo
  {journal} {Phys. Rev. B}\ }\textbf {\bibinfo {volume} {88}},\ \bibinfo
  {pages} {115137} (\bibinfo {year} {2013})}\BibitemShut {NoStop}%
\bibitem [{\citenamefont {Wang}\ and\ \citenamefont
  {Senthil}(2015{\natexlab{a}})}]{WangSenthil2015}%
  \BibitemOpen
  \bibfield  {author} {\bibinfo {author} {\bibfnamefont {C.}~\bibnamefont
  {Wang}}\ and\ \bibinfo {author} {\bibfnamefont {T.}~\bibnamefont {Senthil}},\
  }\href {\doibase 10.1103/PhysRevX.5.041031} {\bibfield  {journal} {\bibinfo
  {journal} {Phys. Rev. X}\ }\textbf {\bibinfo {volume} {5}},\ \bibinfo {pages}
  {041031} (\bibinfo {year} {2015}{\natexlab{a}})}\BibitemShut {NoStop}%
\bibitem [{\citenamefont {Metlitski}\ and\ \citenamefont
  {Vishwanath}(2015)}]{MetlitskiVishwanath2015}%
  \BibitemOpen
  \bibfield  {author} {\bibinfo {author} {\bibfnamefont {M.~A.}\ \bibnamefont
  {Metlitski}}\ and\ \bibinfo {author} {\bibfnamefont {A.}~\bibnamefont
  {Vishwanath}},\ }\href@noop {} {\enquote {\bibinfo {title} {Particle-vortex
  duality of 2d {Dirac} fermion from electric-magnetic duality of 3d
  topological insulators},}\ } (\bibinfo {year} {2015}),\ \bibinfo {note}
  {unpublished},\ \Eprint {http://arxiv.org/abs/1505.05142} {arXiv:1505.05142
  [cond-mat.str-el]} \BibitemShut {NoStop}%
\bibitem [{\citenamefont {Wang}\ and\ \citenamefont
  {Senthil}(2015{\natexlab{b}})}]{WangSenthilReview}%
  \BibitemOpen
  \bibfield  {author} {\bibinfo {author} {\bibfnamefont {C.}~\bibnamefont
  {Wang}}\ and\ \bibinfo {author} {\bibfnamefont {T.}~\bibnamefont {Senthil}},\
  }\href@noop {} {\enquote {\bibinfo {title} {Half-filled {Landau} level,
  topological insulator surfaces, and three dimensional quantum spin
  liquids},}\ } (\bibinfo {year} {2015}{\natexlab{b}}),\ \bibinfo {note}
  {unpublished},\ \Eprint {http://arxiv.org/abs/1507.08290} {arXiv:1507.08290}
  \BibitemShut {NoStop}%
\bibitem [{Note1()}]{Note1}%
  \BibitemOpen
  \bibinfo {note} {Such a particle-hole symmetric Dirac theory can arise at the
  surface of a class AIII 3D topological superconductor \cite
  {MetlitskiVishwanath2015,WangSenthilReview}.}\BibitemShut {Stop}%
\bibitem [{\citenamefont {Son}(2015)}]{Son}%
  \BibitemOpen
  \bibfield  {author} {\bibinfo {author} {\bibfnamefont {D.~T.}\ \bibnamefont
  {Son}},\ }\href {\doibase 10.1103/PhysRevX.5.031027} {\bibfield  {journal}
  {\bibinfo  {journal} {Phys. Rev. X}\ }\textbf {\bibinfo {volume} {5}},\
  \bibinfo {pages} {031027} (\bibinfo {year} {2015})}\BibitemShut {NoStop}%
\bibitem [{\citenamefont {Geraedts}\ \emph {et~al.}(2015)\citenamefont
  {Geraedts}, \citenamefont {Zaletel}, \citenamefont {Mong}, \citenamefont
  {Metlitski}, \citenamefont {Vishwanath},\ and\ \citenamefont
  {Motrunich}}]{Geraedts}%
  \BibitemOpen
  \bibfield  {author} {\bibinfo {author} {\bibfnamefont {S.~D.}\ \bibnamefont
  {Geraedts}}, \bibinfo {author} {\bibfnamefont {M.~P.}\ \bibnamefont
  {Zaletel}}, \bibinfo {author} {\bibfnamefont {R.~S.~K.}\ \bibnamefont
  {Mong}}, \bibinfo {author} {\bibfnamefont {M.~A.}\ \bibnamefont {Metlitski}},
  \bibinfo {author} {\bibfnamefont {A.}~\bibnamefont {Vishwanath}}, \ and\
  \bibinfo {author} {\bibfnamefont {O.~I.}\ \bibnamefont {Motrunich}},\
  }\href@noop {} {\enquote {\bibinfo {title} {The half-filled {Landau} level:
  the case for {Dirac} composite fermions},}\ } (\bibinfo {year} {2015}),\
  \bibinfo {note} {unpublished},\ \Eprint {http://arxiv.org/abs/1508.04140}
  {arXiv:1508.04140} \BibitemShut {NoStop}%
\bibitem [{\citenamefont {Murthy}\ and\ \citenamefont
  {Shankar}(2015)}]{ShankarMurthy}%
  \BibitemOpen
  \bibfield  {author} {\bibinfo {author} {\bibfnamefont {G.}~\bibnamefont
  {Murthy}}\ and\ \bibinfo {author} {\bibfnamefont {R.}~\bibnamefont
  {Shankar}},\ }\href@noop {} {\enquote {\bibinfo {title} {The
  {\ensuremath$\nu=1/2$} {Landau} level: Half-full or half-empty?}}\ }
  (\bibinfo {year} {2015}),\ \bibinfo {note} {unpublished},\ \Eprint
  {http://arxiv.org/abs/1508.06974} {arXiv:1508.06974} \BibitemShut {NoStop}%
\bibitem [{\citenamefont {Kachru}\ \emph {et~al.}(2015)\citenamefont {Kachru},
  \citenamefont {Mulligan}, \citenamefont {Torroba},\ and\ \citenamefont
  {Wang}}]{Kachru}%
  \BibitemOpen
  \bibfield  {author} {\bibinfo {author} {\bibfnamefont {S.}~\bibnamefont
  {Kachru}}, \bibinfo {author} {\bibfnamefont {M.}~\bibnamefont {Mulligan}},
  \bibinfo {author} {\bibfnamefont {G.}~\bibnamefont {Torroba}}, \ and\
  \bibinfo {author} {\bibfnamefont {H.}~\bibnamefont {Wang}},\ }\href {\doibase
  10.1103/PhysRevB.92.235105} {\bibfield  {journal} {\bibinfo  {journal} {Phys.
  Rev. B}\ }\textbf {\bibinfo {volume} {92}},\ \bibinfo {pages} {235105}
  (\bibinfo {year} {2015})}\BibitemShut {NoStop}%
\bibitem [{\citenamefont {{Metlitski}}(2015)}]{metlitskiduality}%
  \BibitemOpen
  \bibfield  {author} {\bibinfo {author} {\bibfnamefont {M.~A.}\ \bibnamefont
  {{Metlitski}}},\ }\href@noop {} {\enquote {\bibinfo {title}
  {{\ensuremath{S}-duality of \ensuremath{u(1)} gauge theory with
  \ensuremath{\theta =\pi} on non-orientable manifolds: Applications to
  topological insulators and superconductors}},}\ } (\bibinfo {year} {2015}),\
  \Eprint {http://arxiv.org/abs/1510.05663} {arXiv:1510.05663 [hep-th]}
  \BibitemShut {NoStop}%
\bibitem [{\citenamefont {Vishwanath}\ and\ \citenamefont
  {Senthil}(2013)}]{VishwanathSenthil2013}%
  \BibitemOpen
  \bibfield  {author} {\bibinfo {author} {\bibfnamefont {A.}~\bibnamefont
  {Vishwanath}}\ and\ \bibinfo {author} {\bibfnamefont {T.}~\bibnamefont
  {Senthil}},\ }\href {\doibase 10.1103/PhysRevX.3.011016} {\bibfield
  {journal} {\bibinfo  {journal} {Phys. Rev. X}\ }\textbf {\bibinfo {volume}
  {3}} (\bibinfo {year} {2013}),\ 10.1103/PhysRevX.3.011016}\BibitemShut
  {NoStop}%
\bibitem [{\citenamefont {Wen}(1991)}]{XGanomaly}%
  \BibitemOpen
  \bibfield  {author} {\bibinfo {author} {\bibfnamefont {X.~G.}\ \bibnamefont
  {Wen}},\ }\href {\doibase 10.1103/PhysRevB.43.11025} {\bibfield  {journal}
  {\bibinfo  {journal} {Phys. Rev. B}\ }\textbf {\bibinfo {volume} {43}},\
  \bibinfo {pages} {11025} (\bibinfo {year} {1991})}\BibitemShut {NoStop}%
\bibitem [{\citenamefont {Mong}\ \emph {et~al.}(2010)\citenamefont {Mong},
  \citenamefont {Essin},\ and\ \citenamefont {Moore}}]{AFTImong}%
  \BibitemOpen
  \bibfield  {author} {\bibinfo {author} {\bibfnamefont {R.~S.~K.}\
  \bibnamefont {Mong}}, \bibinfo {author} {\bibfnamefont {A.~M.}\ \bibnamefont
  {Essin}}, \ and\ \bibinfo {author} {\bibfnamefont {J.~E.}\ \bibnamefont
  {Moore}},\ }\href {\doibase 10.1103/PhysRevB.81.245209} {\bibfield  {journal}
  {\bibinfo  {journal} {Phys. Rev. B}\ }\textbf {\bibinfo {volume} {81}},\
  \bibinfo {pages} {245209} (\bibinfo {year} {2010})}\BibitemShut {NoStop}%
\bibitem [{\citenamefont {Essin}\ and\ \citenamefont
  {Gurarie}(2012)}]{AFTIessin}%
  \BibitemOpen
  \bibfield  {author} {\bibinfo {author} {\bibfnamefont {A.~M.}\ \bibnamefont
  {Essin}}\ and\ \bibinfo {author} {\bibfnamefont {V.}~\bibnamefont
  {Gurarie}},\ }\href {\doibase 10.1103/PhysRevB.85.195116} {\bibfield
  {journal} {\bibinfo  {journal} {Phys. Rev. B}\ }\textbf {\bibinfo {volume}
  {85}},\ \bibinfo {pages} {195116} (\bibinfo {year} {2012})}\BibitemShut
  {NoStop}%
\bibitem [{\citenamefont {Fang}\ \emph {et~al.}(2013)\citenamefont {Fang},
  \citenamefont {Gilbert},\ and\ \citenamefont {Bernevig}}]{AFTIfang}%
  \BibitemOpen
  \bibfield  {author} {\bibinfo {author} {\bibfnamefont {C.}~\bibnamefont
  {Fang}}, \bibinfo {author} {\bibfnamefont {M.~J.}\ \bibnamefont {Gilbert}}, \
  and\ \bibinfo {author} {\bibfnamefont {B.~A.}\ \bibnamefont {Bernevig}},\
  }\href {\doibase 10.1103/PhysRevB.88.085406} {\bibfield  {journal} {\bibinfo
  {journal} {Phys. Rev. B}\ }\textbf {\bibinfo {volume} {88}},\ \bibinfo
  {pages} {085406} (\bibinfo {year} {2013})}\BibitemShut {NoStop}%
\bibitem [{\citenamefont {Dasgupta}\ and\ \citenamefont
  {Halperin}(1981)}]{DasguptaHalperin}%
  \BibitemOpen
  \bibfield  {author} {\bibinfo {author} {\bibfnamefont {C.}~\bibnamefont
  {Dasgupta}}\ and\ \bibinfo {author} {\bibfnamefont {B.~I.}\ \bibnamefont
  {Halperin}},\ }\href {\doibase 10.1103/PhysRevLett.47.1556} {\bibfield
  {journal} {\bibinfo  {journal} {Phys. Rev. Lett.}\ }\textbf {\bibinfo
  {volume} {47}},\ \bibinfo {pages} {1556} (\bibinfo {year}
  {1981})}\BibitemShut {NoStop}%
\bibitem [{\citenamefont {Fisher}\ and\ \citenamefont
  {Lee}(1989)}]{MatthewDungHai}%
  \BibitemOpen
  \bibfield  {author} {\bibinfo {author} {\bibfnamefont {M.~P.~A.}\
  \bibnamefont {Fisher}}\ and\ \bibinfo {author} {\bibfnamefont {D.~H.}\
  \bibnamefont {Lee}},\ }\href {\doibase 10.1103/PhysRevB.39.2756} {\bibfield
  {journal} {\bibinfo  {journal} {Phys. Rev. B}\ }\textbf {\bibinfo {volume}
  {39}},\ \bibinfo {pages} {2756} (\bibinfo {year} {1989})}\BibitemShut
  {NoStop}%
\bibitem [{\citenamefont {Fu}\ and\ \citenamefont {Kane}(2008)}]{FuKane}%
  \BibitemOpen
  \bibfield  {author} {\bibinfo {author} {\bibfnamefont {L.}~\bibnamefont
  {Fu}}\ and\ \bibinfo {author} {\bibfnamefont {C.~L.}\ \bibnamefont {Kane}},\
  }\href {\doibase 10.1103/PhysRevLett.100.096407} {\bibfield  {journal}
  {\bibinfo  {journal} {Phys. Rev. Lett.}\ }\textbf {\bibinfo {volume} {100}},\
  \bibinfo {pages} {096407} (\bibinfo {year} {2008})}\BibitemShut {NoStop}%
\bibitem [{\citenamefont {Mross}\ \emph {et~al.}(2015)\citenamefont {Mross},
  \citenamefont {Essin},\ and\ \citenamefont {Alicea}}]{STO}%
  \BibitemOpen
  \bibfield  {author} {\bibinfo {author} {\bibfnamefont {D.~F.}\ \bibnamefont
  {Mross}}, \bibinfo {author} {\bibfnamefont {A.}~\bibnamefont {Essin}}, \ and\
  \bibinfo {author} {\bibfnamefont {J.}~\bibnamefont {Alicea}},\ }\href
  {\doibase 10.1103/PhysRevX.5.011011} {\bibfield  {journal} {\bibinfo
  {journal} {Phys. Rev. X}\ }\textbf {\bibinfo {volume} {5}},\ \bibinfo {pages}
  {011011} (\bibinfo {year} {2015})}\BibitemShut {NoStop}%
\bibitem [{\citenamefont {Hermele}\ \emph {et~al.}(2007)\citenamefont
  {Hermele}, \citenamefont {Senthil},\ and\ \citenamefont
  {Fisher}}]{Hermele_ASL_Erratum_2007}%
  \BibitemOpen
  \bibfield  {author} {\bibinfo {author} {\bibfnamefont {M.}~\bibnamefont
  {Hermele}}, \bibinfo {author} {\bibfnamefont {T.}~\bibnamefont {Senthil}}, \
  and\ \bibinfo {author} {\bibfnamefont {M.~P.~A.}\ \bibnamefont {Fisher}},\
  }\href {\doibase 10.1103/PhysRevB.76.149906} {\bibfield  {journal} {\bibinfo
  {journal} {Phys. Rev. B}\ }\textbf {\bibinfo {volume} {76}},\ \bibinfo
  {pages} {149906} (\bibinfo {year} {2007})}\BibitemShut {NoStop}%
\bibitem [{Note2()}]{Note2}%
  \BibitemOpen
  \bibinfo {note} {In fact, $N=2$ QED$_3$ exhibits additional symmetries whose
  action on the dual bosons $b_{\sigma ,y}$ is derived in the
  Appendix.}\BibitemShut {Stop}%
\bibitem [{\citenamefont {{Wang}}\ and\ \citenamefont
  {{Senthil}}(2015)}]{wangsenthilu1}%
  \BibitemOpen
  \bibfield  {author} {\bibinfo {author} {\bibfnamefont {C.}~\bibnamefont
  {{Wang}}}\ and\ \bibinfo {author} {\bibfnamefont {T.}~\bibnamefont
  {{Senthil}}},\ }\href@noop {} {\enquote {\bibinfo {title} {{Time-reversal
  symmetric U(1) quantum spin liquids}},}\ } (\bibinfo {year} {2015}),\ \Eprint
  {http://arxiv.org/abs/1505.03520} {arXiv:1505.03520 [cond-mat.str-el]}
  \BibitemShut {NoStop}%
\bibitem [{\citenamefont {Xu}\ and\ \citenamefont {You}(2015)}]{XuYouSPT}%
  \BibitemOpen
  \bibfield  {author} {\bibinfo {author} {\bibfnamefont {C.}~\bibnamefont
  {Xu}}\ and\ \bibinfo {author} {\bibfnamefont {Y.-Z.}\ \bibnamefont {You}},\
  }\href@noop {} {\enquote {\bibinfo {title} {Self-dual quantum electrodynamics
  as boundary state of the three dimensional bosonic topological insulator},}\
  } (\bibinfo {year} {2015}),\ \bibinfo {note} {unpublished},\ \Eprint
  {http://arxiv.org/abs/1510.06032} {arXiv:1510.06032} \BibitemShut {NoStop}%
\bibitem [{\citenamefont {Murthy}\ and\ \citenamefont
  {Shankar}(2003)}]{MurthyShankarRMP}%
  \BibitemOpen
  \bibfield  {author} {\bibinfo {author} {\bibfnamefont {G.}~\bibnamefont
  {Murthy}}\ and\ \bibinfo {author} {\bibfnamefont {R.}~\bibnamefont
  {Shankar}},\ }\href {\doibase 10.1103/RevModPhys.75.1101} {\bibfield
  {journal} {\bibinfo  {journal} {Rev. Mod. Phys.}\ }\textbf {\bibinfo {volume}
  {75}},\ \bibinfo {pages} {1101} (\bibinfo {year} {2003})}\BibitemShut
  {NoStop}%
\bibitem [{\citenamefont {Herzog}\ \emph {et~al.}(2007)\citenamefont {Herzog},
  \citenamefont {Kovtun}, \citenamefont {Sachdev},\ and\ \citenamefont
  {Son}}]{Herzog2007}%
  \BibitemOpen
  \bibfield  {author} {\bibinfo {author} {\bibfnamefont {C.~P.}\ \bibnamefont
  {Herzog}}, \bibinfo {author} {\bibfnamefont {P.}~\bibnamefont {Kovtun}},
  \bibinfo {author} {\bibfnamefont {S.}~\bibnamefont {Sachdev}}, \ and\
  \bibinfo {author} {\bibfnamefont {D.~T.}\ \bibnamefont {Son}},\ }\href
  {\doibase 10.1103/PhysRevD.75.085020} {\bibfield  {journal} {\bibinfo
  {journal} {Phys. Rev. D}\ }\textbf {\bibinfo {volume} {75}},\ \bibinfo
  {pages} {085020} (\bibinfo {year} {2007})}\BibitemShut {NoStop}%
\bibitem [{\citenamefont {Geraedts}\ and\ \citenamefont
  {Motrunich}(2012)}]{Geraedts_rangedloops}%
  \BibitemOpen
  \bibfield  {author} {\bibinfo {author} {\bibfnamefont {S.~D.}\ \bibnamefont
  {Geraedts}}\ and\ \bibinfo {author} {\bibfnamefont {O.~I.}\ \bibnamefont
  {Motrunich}},\ }\href {\doibase 10.1103/PhysRevB.85.144303} {\bibfield
  {journal} {\bibinfo  {journal} {Phys. Rev. B}\ }\textbf {\bibinfo {volume}
  {85}},\ \bibinfo {pages} {144303} (\bibinfo {year} {2012})}\BibitemShut
  {NoStop}%
\bibitem [{\citenamefont {Haldane}(1988)}]{Haldanemodel}%
  \BibitemOpen
  \bibfield  {author} {\bibinfo {author} {\bibfnamefont {F.~D.~M.}\
  \bibnamefont {Haldane}},\ }\href {\doibase 10.1103/PhysRevLett.61.2015}
  {\bibfield  {journal} {\bibinfo  {journal} {Phys. Rev. Lett.}\ }\textbf
  {\bibinfo {volume} {61}},\ \bibinfo {pages} {2015} (\bibinfo {year}
  {1988})}\BibitemShut {NoStop}%
\bibitem [{\citenamefont {Chalker}\ and\ \citenamefont
  {Coddington}(1988)}]{Chalker}%
  \BibitemOpen
  \bibfield  {author} {\bibinfo {author} {\bibfnamefont {J.~T.}\ \bibnamefont
  {Chalker}}\ and\ \bibinfo {author} {\bibfnamefont {P.~D.}\ \bibnamefont
  {Coddington}},\ }\href@noop {} {\bibfield  {journal} {\bibinfo  {journal} {J.
  Phys. C}\ }\textbf {\bibinfo {volume} {21}},\ \bibinfo {pages} {2665}
  (\bibinfo {year} {1988})}\BibitemShut {NoStop}%
\bibitem [{\citenamefont {Kane}\ \emph {et~al.}(2002)\citenamefont {Kane},
  \citenamefont {Mukhopadhyay},\ and\ \citenamefont {Lubensky}}]{KaneWires}%
  \BibitemOpen
  \bibfield  {author} {\bibinfo {author} {\bibfnamefont {C.~L.}\ \bibnamefont
  {Kane}}, \bibinfo {author} {\bibfnamefont {R.}~\bibnamefont {Mukhopadhyay}},
  \ and\ \bibinfo {author} {\bibfnamefont {T.~C.}\ \bibnamefont {Lubensky}},\
  }\href {\doibase 10.1103/PhysRevLett.88.036401} {\bibfield  {journal}
  {\bibinfo  {journal} {Phys. Rev. Lett.}\ }\textbf {\bibinfo {volume} {88}},\
  \bibinfo {pages} {036401} (\bibinfo {year} {2002})}\BibitemShut {NoStop}%
\end{thebibliography}%

\onecolumngrid
\appendix
\section{\large Appendix}
\section{Discrete symmetries for single Dirac cone/$N = 1$ QED$_3$ dual theories}

We defined the action of the antiunitary symmetries ${\cal T}$ and ${\cal C}$ on the chiral electrons $\psi_y = \eta_y e^{i \phi_y}$ as
\begin{align}
&{\cal T} \psi_y(x) {\cal T}^{-1} = (-1)^y \psi_{y+1}(x) ~,\\
&{\cal C} \psi_y(x) {\cal C}^{-1} = \psi^\dagger_{y+1}(x) ~.
\end{align}
The wire model is additionally symmetric under inversion ${\cal I}$ relative to a point half-way between neighboring wires and mirror ${\cal M}$ defined by
\begin{align}
&{\cal I} \psi_y(x) {\cal I}^{-1} = \psi_{-y+1}(-x) ~,\\
&{\cal M} \psi_y(x) {\cal M}^{-1} = (-1)^y \psi_{-y}(x) ~.
\end{align}
Using a convention where Klein factors transform as ${\cal T} \eta_y {\cal T}^{-1} = {\cal C} \eta_y {\cal C}^{-1} = \eta_{y+1}$, as well as ${\cal I} \eta_y {\cal I}^{-1} = \eta_{-y+1}$ and ${\cal M} \eta_y {\cal M}^{-1} = \eta_{-y}$ , we conclude that the chiral boson fields transform according to
\begin{align}
& {\cal T} \phi_y(x) {\cal T}^{-1} = -\phi_{y+1}(x) + \alpha_y ~, \\
& {\cal C} \phi_y(x) {\cal C}^{-1} = \phi_{y+1}(x) ~, \\ 
& {\cal I} \phi_y(x) {\cal I}^{-1} = \phi_{-y+1}(-x) ~, \\
& {\cal M} \phi_y(x) {\cal M}^{-1} = \phi_{-y}(x) + \alpha_y ~,
\end{align}
where $\alpha_y = y \pi$. The dual modes $\tilde{\phi}_y$ defined via
\begin{equation}
\tilde{\phi}_y(x) \equiv \sum_{y'\neq y} \text{sgn}\(y-y'\) (-1)^{y'} \phi_{y'}(x)
\end{equation}
therefore transform as
\begin{align}
& {\cal T} \tilde{\phi}_{y}(x) {\cal T}^{-1} = \sum_{y'\neq y}\text{sgn}\(y-y'\) (-1)^{y'}\(-\phi_{y'+1}(x)+\alpha_{y'}\) = \tilde{\phi}_{y+1}(x) + \alpha_y + \sum_{y'}\alpha_{y'} ~,\label{phi1}\\
& {\cal C} \tilde{\phi}_y(x) {\cal C}^{-1} = \sum_{y'\neq y}\text{sgn}\(y-y'\) (-1)^{y'}\phi_{y'+1}(x) = -\tilde{\phi}_{y+1}(x) ~,\\ 
& {\cal I} \tilde{\phi}_y(x) {\cal I}^{-1} = \sum_{y'\neq y}\text{sgn}\(y-y'\) (-1)^{y'}\phi_{-y'+1}(-x) = \tilde{\phi}_{-y+1}(-x) ~,\\
& {\cal M} \tilde{\phi}_y(x) {\cal M}^{-1} = \sum_{y'\neq y}\text{sgn}\(y-y'\) (-1)^{y'}(\phi_{-y'}(x) + \alpha_{y'}) = -\tilde{\phi}_{-y}(x)+ \alpha_y + \sum_{y'}\alpha_{y'}~.\label{phi4}
\end{align}
In the first and last equations we used that the phase fields are defined modulo $2\pi$ and $\alpha_{y'} = -\alpha_{y'} \mod 2\pi$ for any $y'$; we will henceforth omit the unimportant constant $\sum_{y'}\alpha_{y'}$. Table \ref{tab.symm1} provides a full list of symmetry transformations for all involved fields.

\begin{table}[h]
\caption{Action of discrete symmetries for single Dirac cone/$N=1$ QED$_3$ duality.}
\begin{tabular}{| c| c| c| c | c| c|c|}
\hline
 &${\Psi}(x, Y)$ & $ \psi_{y}(x)$& $\phi_{y}(x)$& $\tilde \phi_{y}(x)$ &$\tilde\psi_{y}(x)$& $\tilde{\Psi}(x, Y)$\\ \hline
 ${\cal T} \ldots{\cal T}^{-1} $ &$i \sigma^y{\Psi}(x, Y)$&$(-1)^y \psi_{y+1}(x)$&$-\phi_{y+1}(x) + \alpha_y$&$\tilde{\phi}_{y+1}(x) + \alpha_y$&$-\tilde{\psi}_{y+1}^\dagger(x)$&$\sigma^x\tilde{\Psi}(x, Y)$\\
 \hline
 ${\cal C} \ldots{\cal C}^{-1}$ &$\sigma^x{\Psi}^\dagger(x, Y)$&$ \psi^\dagger_{y+1}(x)$&$\phi_{y+1}(x)$&$-\tilde{\phi}_{y+1}(x)$&$-(-1)^y \tilde{\psi}_{y+1}(x)$&$i \sigma^y\tilde{\Psi}^\dagger(x, Y)$\\
 \hline
 ${\cal I} \ldots{\cal I}^{-1}$ &$\sigma^x{\Psi}(-x, -Y)$&$\psi_{-y+1}(-x)$&$\phi_{-y+1}(-x)$&$\tilde{\phi}_{-y+1}(-x)$&$-\tilde{\psi}_{-y+1}(-x)$&$\sigma^x\tilde{\Psi}(-x, -Y)$\\
 \hline
 ${\cal M} \ldots{\cal M}^{-1}$ &$\sigma^z{\Psi}(x, -Y)$&$(-1)^y \psi_{-y}(x)$&$ \phi_{-y}(x) + \alpha_y$&$-\tilde{\phi}_{-y}(x)+ \alpha_y $&$\tilde{\psi}_{-y}^\dagger(x)$&$\sigma^z\tilde{\Psi}^\dagger(x, -Y)$\\
 \hline
\end{tabular}\vspace{.5cm}
\label{tab.symm1}
\end{table}

%%%%%%%%%%%%%%%%%%%%%%%%%%%%%%%%%%%%%%%%%%%%%%%%%%%%%%%%%%%%%%%%%%%%%%%%
\section{Gauge invariance}
\label{app.gauge}
Here we explicitly demonstrate gauge invariance of the bosonized wire model with staggered Chern-Simons term. Consider the action ${\cal S} = {\cal S}_\text{wire} + {\cal S}'_\text{staggered-CS}$
\small
\begin{eqnarray}
{\cal S}_\text{wire} &=& \sum_y \int_{t,x} \left\{\frac{(-1)^y}{4\pi} \partial_x \phi_y \(\partial_t \phi_y - 2 A_{0,y}\) - \frac{v}{4\pi}(\partial_x \phi_y - A_{1,y})^2 \right\} ~,\\
{\cal S}'_\text{staggered-CS} &=& -\sum_y \frac{\nu_y}{4\pi} \int_{t,x} \left\{\Delta A_{0,y} \(A_{1,y+1} + A_{1,y}\)
- \partial_t A_{2, y+1/2} \(A_{1,y+1} + A_{1,y}\) + \partial_x A_{2, y+1/2} \(A_{0,y+1} + A_{0,y}\) \right\} ~,
\end{eqnarray}
\normalsize
where $\Delta A_{\mu,y}=A_{\mu,y+1} - A_{\mu,y}$ as before. One can view ${\cal S}'_\text{staggered-CS}$ as a lattice analogue of Eq.~\eqref{eq.lprime}; for now we leave the coefficient $\nu_y$ general but will fix it momentarily by imposing gauge invariance. Note also that here we do not commit to the $A_2 = 0$ gauge (in contrast to the main text) and label the location of $A_2$ by the midpoint $y+1/2$ between two wires $y$ and $y+1$. 

Suppose that we now perform the gauge transformation
\begin{align}
 &\phi_y \rightarrow \phi_y + f_y ~, \\
 &A_{0,y} \rightarrow A_{0,y} + \partial_t f_y ~,\\
 &A_{1,y} \rightarrow A_{1,y} + \partial_x f_y ~, \\
 &A_{2,y+1/2} \rightarrow A_{2,y+1/2} +\Delta f_y~.
\end{align}
This transformation yields the following anomalies in the separate ${\cal S}_\text{wire}$ and ${\cal S}'_\text{staggered-CS}$ parts:
\begin{align}
 \delta{\cal S}_\text{wire} &= \sum_y \int_{t,x}\frac{-(-1)^y}{4\pi} \left[2 A_{0,y} \partial_x f_y + \partial_t f_y \partial_x f_y \right] ~,\\
 \delta{\cal S}'_\text{staggered-CS} &= \sum_y \int_{t,x} \frac{\nu_{y} - \nu_{y-1}}{4\pi}\left[ 2A_{0,y} \partial_x f_y + \partial_t f_y \partial_x f_y \right] ~.
\end{align}
The total action is gauge invariant when $\delta{\cal S}_\text{wire} + \delta{\cal S}'_\text{staggered-CS} = 0$, which is satisfied for
\begin{align}
\nu_{y}- \nu_{y-1} = (-1)^y \ \ \ \ \Rightarrow \ \ \ \ \ 
\nu_y = \nu_0 + \frac{1}{2}(-1)^y ~.
\end{align}
On the topological insulator surface, the uniform contribution must vanish ($\nu_0 = 0$) by ${\cal T}$ symmetry; we assume that this is the case for the remainder of this Appendix.

%%%%%%%%%%%%%%%%%%%%%%%%%%%%%%%%%%%%%%%%%%%%%%%%%%%%%%%%%%%%%%%%%%%%%% 
\section{Explicit duality in the presence of non-zero electromagnetic vector potential $A_\mu$}

In the main text, we introduced the operators $D$ and $\Delta$ whose matrix representation is
\begin{align}
& D_{y,y'} = \(1-\delta_{y,y'}\)\text{sgn}(y-y')(-1)^{y'} ~,\\
& \Delta_{y,y'} = \delta_{y+1,y'} - \delta_{y,y'} ~.
\end{align}
It is convenient to further define
\begin{align}
&S_{y,y'} = \delta_{y+1,y'} + \delta_{y,y'} ~,\\
& P_{y,y'} = (-1)^y \delta_{y,y'} ~,
\end{align}
which satisfy the following relations:
\begin{eqnarray}
&D P D^T = - P~, \\
&\Delta D = - P \Delta~, \label{relations.deltadp}\\
&S^T \Delta = - \Delta^T S~,\\
&P S P = - \Delta~. \label{relations.deltas}
\end{eqnarray}
The first of these encodes the reversed chirality of $\tilde \psi_y$ compared to $\psi_y$ while the second one is responsible for the locality of dual-fermion hopping when expressed in terms of electrons. The third relation is a lattice analogue of integration by parts. These matrices allow us to express ${\cal S}_\text{wire}$ and ${\cal S}'_\text{staggered-CS}$ concisely (choosing the gauge $A_2=0$ for convenience) as
\begin{eqnarray}
{\cal S}_\text{wire} &=& \int_{t,x} \left[ \frac{P \partial_x \phi \(\partial_t \phi - 2 A_{0}\)}{4\pi} - \frac{v}{4\pi} (\partial_x \phi - A_{1})^2 - h_\text{hop}[\phi] \right] ~,\label{eqn.matrixs1}\\
{\cal S}'_\text{staggered-CS} &=& - \int_{t,x}\frac{P}{8\pi} \Delta A_{0} S A_{1} 
 ~,\label{eqn.matrixs2}
\end{eqnarray}
where $\phi = (\ldots, \phi_{y-1}, \phi_{y}, \phi_{y+1}, \ldots)$, $A_{\mu} = (\ldots, A_{\mu,y-1}, A_{\mu,y}, A_{\mu,y+1}, \ldots)$ and a scalar product is implied, i.e.,\begin{align}B \phi C\phi = \sum_y (B\phi)_y (C\phi)_y = \sum_{y,y',y''} B_{yy'} \phi_{y'} C_{yy''}\phi_{y''} = \phi B^T C \phi\end{align} for matrices $B$,$C$.
The main text derived the duality with $A_\mu=0$ by mapping the electron action
\begin{align}
{\cal S}[A_\mu=0] &= \int_{t,x} \left[\frac{P }{4\pi}\partial_x \phi \partial_t \phi - \frac{v}{4\pi}(\partial_x \phi)^2 - h_\text{hop}[\phi]\right]
\end{align}
onto the dual action
\begin{align}
{\cal S}_\text{dual}[A_\mu=0] &= \int_{t,x} \left[ \frac{-P}{4\pi} \partial_x \tilde{\phi} \(\partial_t \tilde{\phi} - 2 a_{0}\)-\frac{u}{4\pi}(\partial_x \tilde{\phi} - a_1)^2 - h_\text{hop}[\tilde{\phi}] + \frac{1}{16 \pi v}(\Delta a_0)^2 - \frac{v}{16\pi}(\Delta a_1)^2 + \frac{P}{8\pi} \Delta a_0 S a_1 \right] ~.
\end{align}
When restoring the external vector potential, the free-electron action must be supplemented by the $A_\mu$-dependent terms of Eqs.~\eqref{eqn.matrixs1} and \eqref{eqn.matrixs2}. Duality then maps
\begin{align}
{\cal S}={\cal S}[A_\mu =0] &+\int_{t,x}\left[ -\frac{\partial_x \phi P A_0}{2\pi} + \frac{v \partial_x \phi A_1}{2\pi} - \frac{v A_1^2}{4\pi} - \frac{P \Delta A_0 S A_1}{8\pi } \right]\\
 \rightarrow {\cal S}_\text{dual}[A_\mu =0] &+\int_{t,x}\left[ -\frac{\partial_x \tilde{\phi} D^T P A_0}{2\pi} + \frac{v \partial_x \tilde{\phi} D^T A_1}{2\pi} - \frac{v A_1^2}{4\pi} - \frac{P \Delta A_0 S A_1}{8\pi } \right] ~.
\end{align}
Shifting
\begin{align}
a_0 \rightarrow a_0 - D A_0 + v D P A_1 ~
\end{align}
decouples $A_\mu$ from $\tilde \phi$. Using $\Delta (v D P A_1 - D A_0) = v S A_1 + P \Delta A_0$, see Eqs.~\eqref{relations.deltas} and \eqref{relations.deltadp}, we find
\small
\begin{align}
{\cal S}_\text{dual}&={\cal S}_\text{dual}[A_\mu =0]+ \int_{t,x}\bigg{[} \frac{\Delta a_0 \Delta (v D P A_1 - D A_0)}{8 \pi v} + \frac{[\Delta (v D P A_1 - D A_0)]^2}{16 \pi v} + \frac{P\Delta (v D P A_1 - D A_0) S a_1}{8 \pi} - \frac{vA_1^2}{4\pi} - \frac{P\Delta A_0 S A_1}{8\pi} \bigg{]} \nonumber\\
&={\cal S}_\text{dual}[A_\mu =0] + {\cal S}_\text{CS}+{\cal S}_\text{staggered-$aA$}+ \int_{t,x}\bigg{[}\frac{1 }{16 \pi v}(\Delta A_0)^2-\frac{v }{16 \pi}( \Delta A_1)^2 \bigg{]}\nonumber~ .
\end{align}
\normalsize
Here ${\cal S}_\text{staggered-$aA$} = \int_{t,x}P\left( \frac{1}{v}\Delta a_0 \Delta A_0 - v \Delta A_1 \Delta a_1 \right)/8 \pi$ drops out in the continuum limit due to the oscillatory factor $P$ while the final term simply renormalizes the dielectric properties of $A_\mu$. The important $A_\mu$ contribution that survives the continuum limit is the mutual Chern-Simons term ${\cal S}_\text{CS}= \int_{t,x}\left(\Delta a_0 S A_1 + \Delta A_0 S a_1\right)/8 \pi$ as quoted in the main text.

%%%%%%%%%%%%%%%%%%%%%%%%%%%%%%%%%%%%%%%%%%%%%%%%%%%%%%%%%%%%%%%%%%%%

\section{Duality between the T-Pfaffian and Fu-Kane superconductor}
\begin{figure}%[ht]
\includegraphics[width=.6\columnwidth]{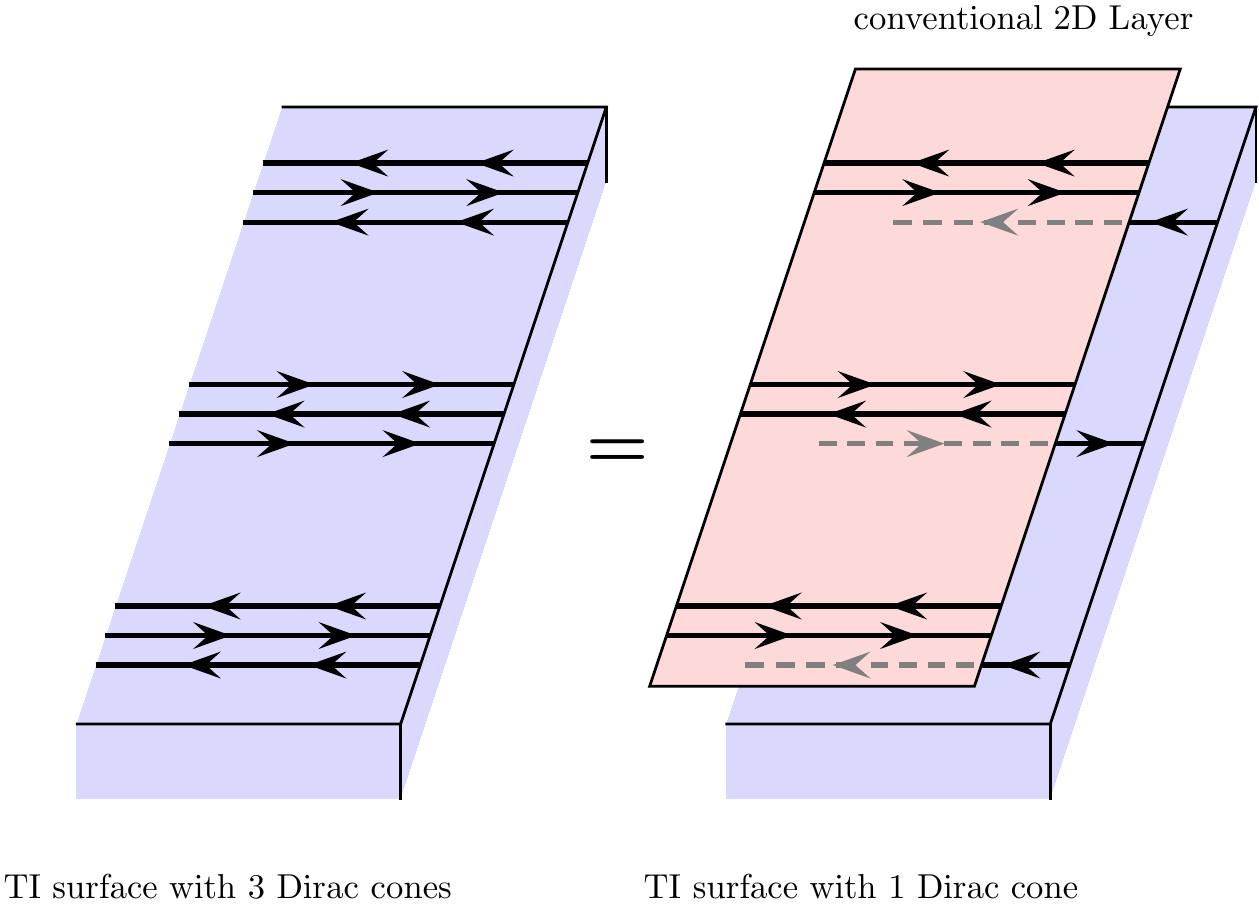}
\caption{Generalized model of the TI surface containing three Dirac cones, only one of which is protected by time-reversal symmetry and charge conservation. As sketched on the right, this setup may equivalently be viewed as a conventional 2D layer containing two Dirac cones, deposited on top of a topological insulator with a single Dirac cone.}
\label{fig:threecone}
\end{figure}

We consider a generalized coupled-wire model where only ${\cal T}^3$ and ${\cal C}^3$ symmetries are enforced, and inter-wire hopping is described by the Hamiltonian density
\begin{align}
 h_\text{hop}=& w_3^{(0)}(-1)^y\,\psi_{3y}^\dagger \psi_{3y+3}+w_3^{(1)}(-1)^y\,\psi_{3y+1}^\dagger \psi_{3y+4}+w_3^{(2)}(-1)^y\,\psi_{3y+2}^\dagger \psi_{3y+5}+{\rm H.c.}~.
\end{align}
These terms only mix electrons between wires whose separation is a multiple of three, and hence give rise to three independent Dirac cones. Two of these may be viewed as forming a strictly two-dimensional system (see Fig.~\ref{fig:threecone}) and can be gapped out by the symmetry-preserving mass term
\begin{align}
 h_\text{mass} = m(-1)^y\, \psi_{3y}^\dagger\psi_{3y+1} + {\rm H.c.},\label{app.mass}
\end{align}
while third Dirac cone remains gapless provided charge is conserved and ${\cal T}^3$ or ${\cal C}^3$ is enforced. 

Instead of being gapped by Eq.~\eqref{app.mass}, the same two Dirac cones can form a conventional 2D superconductor that spontaneously breaks charge conservation when the interaction
\begin{align}
 h_\text{Pair}=& g_\text{Pair}\,\psi_{3y}^\dagger \psi_{3y+1}^\dagger\psi_{3y+3} \psi_{3y+4}+{\rm H.c.}
\end{align}
is relevant. The remaining massless Dirac cone is clearly unaffected by either $h_\text{mass}$ or $h_\text{Pair}$. However, in the latter case the superconductor formed by the first two Dirac cones can proximity-induce pairing into the final Dirac cone via 
\begin{align}
 h_\text{proximity}= g_\text{proximity}\,\psi_{3y+2}^\dagger \psi_{3y+5}^\dagger \psi_{3y+3}\psi_{3y+4}+{\rm H.c.},
\end{align}
without breaking ${\cal T}^3$ or ${\cal C}^3$ symmetry. The inter-wire Hamiltonian density
\begin{align}
h_\text{FK} = h_\text{Pair}+h_\text{proximity}+w_3^{(2)}\,(-1)^y(\psi_{3y+2}^\dagger \psi_{3y+5}+{\rm H.c.})
\end{align}
thus describes a Fu-Kane superconductor \cite{FuKane}. Here the surface electrons are gapped modulo the superfluid mode, present since we are not explicitly breaking charge conservation. One readily verifies its familiar properties, such as Majorana zero-modes trapped in vortex cores.

Consider now the Hamiltonian of a \emph{dual} Fu-Kane superconductor $\tilde h_\text{FK} \equiv h_\text{FK}[\psi_y \rightarrow \tilde \psi_y]$, translated back into the original electron variables $\psi$ using our explicit duality mapping, i.e.,
\begin{align}
 \tilde h_\text{FK} =& g_\text{Pair}\,\psi_{3y}^\dagger\psi_{3y+1}^3\(\psi_{3y+3}^\dagger\)^4\psi_{3y+3}^3 \psi_{3y+4}^\dagger\\
 & +g_\text{proximity}\,\psi_{3y+2}^\dagger \psi_{3y+5}^\dagger \psi_{3y+3}\psi_{3y+4}. \\
 &+ w_3^{(2)}\,(-1)^y\psi_{3y+2}^\dagger(\psi_{3y+3}\psi^\dagger_{3y+4})^2\psi_{3y+5}+{\rm H.c.}.
\end{align}
Precisely this Hamiltonian was derived in Ref.~\onlinecite{STO} for the T-Pfaffian topological order. The $g_\text{Pair}$-term gaps charge fluctuations, while the other two terms gap the remaining neutral degrees of freedom.

%%%%%%%%%%%%%%%%%%%%%%%%%%%%%%%%%%%%%%%%%%%%%%%%%%%%%%%%%%%%%%%%%%%%%

\section{Continuum limit and the dual-fermion velocity}
The dual action for $A_\mu = 0$ is once again given by
\begin{eqnarray}
{\cal S}_\text{dual} = \int_{t,x}\sum_y && \left[\frac{-(-1)^y \partial_x \tilde{\phi}_y \(\partial_t \tilde{\phi}_y - 2 a_{0,y}\) }{4\pi}- \frac{u}{4\pi} \(\partial_x \tilde{\phi}_y - a_{1,y}\)^2 \right. \\
&&~ \left. +\frac{1}{16 \pi v}(\Delta a_{0,y})^2- \frac{v}{16 \pi}(\Delta a_{1,y})^2 + \frac{(-1)^y}{8 \pi} \Delta a_{0,y} (a_{1,y+1} + a_{1,y}) \right] ~.\label{sdual}
\end{eqnarray}
We see that the gauge field is massless only near zero momentum in the $y$-direction, and hence the most naive continuum limit would simply drop the last staggered Chern-Simons term in the action. Here we will pursue a more accurate procedure that formally integrates out such high-momentum massive modes, which yields useful insight into parameters for the dual theory. 

To this end note that the Chern-Simons term couples zero-momentum modes with gapped modes at momentum $\pi$. Hence, we decompose $a_\mu$ into smooth and rapidly oscillatory pieces
\begin{align}
a_{\mu, y} \approx a_{\mu, y}^{(0)} + (-1)^y a_{\mu, y}^{(\pi)} ~,
\end{align}
where both $a_{\mu, y}^{(0)}$ and $a_{\mu, y}^{(\pi)}$ are slowly varying functions of $y$. The action becomes
\begin{eqnarray}
{\cal S}_\text{dual} &\approx&
 \int_{t,x}\sum_y \left[ \frac{-(-1)^y \partial_x \tilde{\phi}_y \(\partial_t \tilde{\phi}_y - 2 a_{0,y}^{(0)}\)}{4\pi} - \frac{u}{4\pi} \(\partial_x \tilde{\phi}_y - a_{1,y}^{(0)} \)^2 - \frac{v}{16\pi}\(\Delta a_{1,y}^{(0)}\)^2 + \frac{1}{16\pi v}\(\Delta a_{0,y}^{(0)}\)^2 \right] \\
&+& \int_{t,x}\sum_y \left[ \frac{a_{0,y}^{(\pi)} \(\partial_x \tilde{\phi}_y - a_{1,y}^{(0)}\)}{2\pi} + \frac{u}{2\pi} \(\partial_x \tilde{\phi}_y - a_{1,y}^{(0)}\) (-1)^y a_{1,y}^{(\pi)} - \frac{u + v}{4\pi}\(a_{1,y}^{(\pi)}\)^2 + \frac{1}{4 \pi v}\(a_{0,y}^{(\pi)}\)^2 - \frac{\Delta a_{0,y}^{(0)} \Delta a_{1,y}^{(\pi)}}{8 \pi} \right] ~, \nonumber
\end{eqnarray}
where we dropped rapidly oscillating terms (note however that we are not making any assumptions about $\tilde{\phi}_y$ and all terms containing these fields are kept). At this point, integrating out the massive $a_{\mu}^{(\pi)}$ modes yields
\begin{equation}
{\cal S}_\text{dual, eff.} \approx
 \int_{t,x}\sum_y \left[\frac{-(-1)^y \partial_x \tilde{\phi}_y \(\partial_t \tilde{\phi}_y - 2 a_{0,y}^{(0)}\) }{4\pi}- \frac{\tilde{v}}{4\pi} \(\partial_x \tilde{\phi}_y - a_{1,y}^{(0)} \)^2 - \frac{v}{16 \pi}\(\Delta a_{1,y}^{(0)}\)^2 + \frac{1}{16 \pi v}\(\Delta a_{0,y}^{(0)}\)^2 \right] ~, \label{sapprox}
\end{equation}
up to terms containing higher derivatives. In this more systematic approach the dual-fermion velocity has been renormalized from $u$ to $\tilde{v} = v + \frac{u v}{u+v}$. We note that while the microscopic theory ${\cal Z} \sim \int e^{i {\cal S}_\text{dual}}$ is independent of $u$ (up to a constant multiplying ${\cal Z}$ and assuming for certainty non-negative $u$), this appears to be no longer the case in Eq.~\eqref{sapprox}. 
This is because keeping only $a_\mu^{(0)}$ and $a_\mu^{(\pi)}$ modes in the above analysis is an approximation; recovering the $u$-independence would demand a more careful treatment of the massive modes, which however would modify only short-range interactions. Universal properties will not depend on such short-range details, and it is sufficient to observe that $\tilde{v} \sim v$ (with weak $u$ dependence). That is, the dual fermions have robust effective kinetic energy along the wires, even if it appears that the bare velocity parameter $u$ was introduced arbitrarily.

%%%%%%%%%%%%%%%%%%%%%%%%%%%%%%%%%%%%%%%%%%%%%%%%%%%%%%%%%%%%%%%%%%%%%%%%%

\section{Stability of $N=1$ lattice QED$_3$ with generic Maxwell term}

Our explicit derivation showed that free Dirac electrons are dual to QED$_3$ where the Maxwell term of the emergent photon 
\begin{align}{\cal S}_\text{MW}\sim \lambda_\mu\left( \epsilon_{\mu\nu\kappa}\partial_\nu a_{\kappa}\right)^2\end{align}
has the bare parameter $\lambda_y=0$. Although a finite $\lambda_y$ is generated under renormalization, it is instructive to add by hand a non-zero bare $\lambda_y$ and discuss its effect on the original Dirac electrons. This is most conveniently done by applying the duality transformation to ${\cal S}_\text{dual}$ of Eq.~\eqref{eqn.sdual} (now with $\lambda_y \neq 0$), resulting in
\begin{align}
{\cal S} &= \int_{t,x}\sum_y\left\{\frac{(-1)^y \partial_x \phi_y \partial_t \phi_y}{4\pi} - h_{\text{wire}}[\phi,a'_\mu] - h_\text{hop}[\phi]\right\}\\
&+ {\cal S}_\text{MW}[a_\mu]\big|_{\lambda_y\neq 0} + {\cal S}_\text{MW}[a'_\mu]\big|_{\lambda_y=0} + {\cal S}_\text{CS}[a_\mu,a'_\mu].
\end{align}
Here $a'_\mu$ is a new emergent gauge field resulting from applying duality a second time, and we suppressed all terms that vanish in the continuum limit. Because of the mutual Chern-Simons term ${\cal S}_\text{CS}$, integrating out $a_\mu$ renders $a'_\mu$ massive. Massive $a'_\mu$ bosons only mediate short-range interactions that are irrelevant at the charge neutrality point and hence do not destabilize the electronic Dirac liquid. Thus, $N=1$ lattice QED$_3$ with general bare $\lambda_y$ maps onto a weakly correlated Dirac liquid of electrons whose short-range interaction strength is determined by $\lambda_y$.

%%%%%%%%%%%%%%%%%%%%%%%%%%%%%%%%%%%%%%%%%%%%%%%%%%%%%%%%%%%%%%%%%%%%%%%%
%%%%%%%%%%%%%%%%%%%%%%%%%%%%%%%%%%%%%%%%%%%%%%%%%%%%%%%%%%%%%%%%%%%%%%%%
\section{More properties of $N = 1$ QED$_3$: Duals of fermion currents}

Here we show another interesting example of relating operators in $N = 1$ QED$_3$ to operators in free Dirac theory. Proceeding similarly to our discussion of the fermion mass term in Eq.~\eqref{massterms}, we can use the microscopic Eq. (12) to also obtain
\begin{equation}
\int_x \sum_y (i \tilde{\psi}_y^\dagger \tilde{\psi}_{y+1} + \text{H.c.}) =\int_x \sum_y (-1)^{y+1} (i \psi_y^\dagger \psi_{y+1} + \text{H.c.}) ~.
\end{equation}
Taking the continuum limit of the two sides gives the identification (omitting numerical factors)
\begin{equation}
\tilde{\Psi}^\dagger \sigma^x i\partial_y \tilde{\Psi}
\sim \Psi^\dagger \sigma^y \Psi \equiv j_2 ~,
\label{j2}
\end{equation}
where $j_2$ is an electron current in the $\hat{y}$ direction. A similar equation with $\tilde{\psi}$ and $\psi$ interchanged gives
\begin{equation}
\Psi^\dagger \sigma^x i\partial_y \Psi \sim \tilde{\Psi}^\dagger \sigma^y \tilde{\Psi} \equiv \tilde{j}_2 ~,
\label{tilde_j2}
\end{equation}
where $\tilde{j}_2$ is a fermion current in the $\hat{y}$ direction in the dual QED$_3$ theory. Thus, we can relate a fermion current in one theory to a fermion bilinear with a derivative in the other theory and vice versa. 
We will see that these unusual identifications are consistent with the expected properties of QED$_3$.

Let us first consider the second relation identifying the current $\tilde{j}_2$ in the dual theory and the specific electron bilinear with one derivative. The scaling dimension of the latter in the free Dirac theory is $3$. This implies that correlations of the $\tilde{j}_2$ currents decay as $1/|\br|^6$, with $\br$ a space-time coordinate, corresponding to a Fourier space non-analyticity $\langle \tilde{j}_2(\bq) \tilde{j}_2(-\bq) \rangle_\text{sing} \sim |\bq|^3$. As we will see, this is expected because the currents $\tilde{j}$ have long-range interactions mediated by the gauge field in the QED$_3$ theory. Indeed, let us remind ourselves how this works in a generic matter-gauge theory (and we will also see an often-stated relation between direct and dual ``conductivities''). Consider a (2+1)D path integral for matter fields $\tilde{\Psi}_{\text{matter}}$ coupled to a gauge field $a_\mu$ 
\begin{equation}
Z[c_\mu] = \int{\cal D}\tilde{\Psi}_{\text{matter}} \,{\cal D}a_\mu \,
\exp\left[-S_{\text{matter}}[\tilde{\Psi}_{\text{matter}}] - \frac{\kappa}{2} \int_\br (\epsilon_{\mu\nu\lambda} \partial_\nu a_\lambda)^2 - i \int_\br \tilde{j}_\mu a_\mu - i \int_\br \tilde{j}_\mu c_\mu \right] ~.
\end{equation}
To simplify equations, in this Appendix only, we work in imaginary time (i.e., Euclidean path integral) and consider a space-time-isotropic Maxwell term for the gauge field and space-time-isotropic full action. We do not need any details about the matter field $\tilde{\Psi}_\text{matter}$ other than the exhibited coupling of its current $\tilde{j}_\mu$ to the dynamical gauge field $a_\mu$. Note that we use tilde over the matter-field objects only to match our earlier notation on the QED$_3$ side of the duality.

When defining the above path integral, we have also included a probing gauge field $c_\mu(\br)$, which allows us to calculate current correlations by functional differentiation:
\begin{equation}
\langle \tilde{j}_\mu(\br) \tilde{j}_{\mu'}({\bm 0}) \rangle = -\frac{1}{Z} \frac{\delta^2 Z}{\delta c_\mu(\br) \delta c_{\mu'}({\bm 0})} \Big{|}_{c_\mu = 0} ~.
\label{delta2Zdcdc}
\end{equation}
Upon shifting the integration variables $a_\mu \to a_\mu - c_\mu$, we can recast the path integral as
\begin{equation}
Z[c_\mu] = \int {\cal D}\tilde{\Psi}_\text{matter} \,{\cal D}a_\mu \,
\exp\left[-S_\text{matter}[\tilde{\Psi}_\text{matter}] - \frac{\kappa}{2} \int_\br (\epsilon_{\mu\nu\lambda} \partial_\nu a_\lambda - \epsilon_{\mu\nu\lambda} \partial_\nu c_\lambda)^2 - i \int_\br \tilde{j}_\mu a_\mu \right] ~.
\end{equation}
By applying Eq.~(\ref{delta2Zdcdc}) to this form, upon differentiation and setting $c_\mu = 0$, we obtain the following relation
\begin{equation}
\langle \tilde{j}_\mu(\br) \tilde{j}_{\mu'}({\bm 0}) \rangle = 
\kappa (\nabla_\mu \nabla_{\mu'} - \delta_{\mu\mu'} {\bm \nabla}^2) \delta(\br)
- \kappa^2 \langle \epsilon_{\mu\nu\lambda} \partial_\nu J_\lambda(\br) \,\,
\epsilon_{\mu'\nu'\lambda'} \partial_{\nu'} J_{\lambda'}({\bm 0}) \rangle ~, 
\label{tildejcorr}
\end{equation}
where $J_\mu(\br) \equiv \epsilon_{\mu\nu\lambda} \partial_\nu a_\lambda(\br)$ denotes the gauge flux. The first term on the right-hand-side is local, so the long-distance behavior is determined by the second term. Specializing to critical field theories, the gauge flux $J_\mu$ in the strongly-coupled CFT has scaling dimension 2: In the $N = 1$ QED$_3$ theory, we know this from the duality since $J_\mu$ gives the original free Dirac electron current (more precisely, $j_\mu = \frac{J_\mu}{4\pi}$);
for general $N$, the scaling dimension is fixed by the fact that $J_\mu$ is a conserved current with short-range interactions and can be also established using large-$N$ treatments. Hence, using Eq.~(\ref{tildejcorr}) we conclude that the $\tilde{j}$ correlations show $1/|\br|^6$ power law decay, in agreement with our discussion of Eq.~(\ref{tilde_j2}). Note, however, that this scaling dimension is in principle already known within the QED$_3$ theory, so in this sense we did not find a qualitatively new property.

Let us consider now Eq.~(\ref{j2}), which relates a specific dual fermion bilinear with a derivative to the electron current $j_2$ in the free Dirac theory. The latter has scaling dimension 2. Hence $\tilde{\Psi}^\dagger \sigma^x i\partial_y \tilde{\Psi}$ in the QED$_3$ theory also has the same scaling dimension, which is reduced (corresponding to enhanced correlations) compared to non-interacting fermions and is a non-trivial effect due to the gauge field in the QED$_3$ theory. However, in retrospect we could have anticipated this just within the QED$_3$ theory, since this operator has the same symmetry transformation properties as the gauge flux $J_2$, which as discussed has scaling dimension $2$. Thus, this is also not a qualitatively new prediction in the QED$_3$ theory, but provides a nice consistency check for the duality.

Continuing a bit more with general considerations, let us also examine the current correlation in Fourier space. For divergenceless currents $\tilde{j}_\mu$ and $J_\mu$, we can write
\begin{equation}
\langle \tilde{j}_\mu(\bq) \tilde{j}_{\mu'}(-\bq) \rangle 
\equiv \cK_{\tilde{j}}(\bq) \left(\delta_{\mu\mu'} - \frac{q_\mu q_{\mu'}}{\bq^2} \right) ~,
\quad\quad
\langle J_\mu(\bq) J_{\mu'}(-\bq) \rangle
\equiv \cK_J(\bq) \left(\delta_{\mu\mu'} - \frac{q_\mu q_{\mu'}}{\bq^2} \right) ~.
\end{equation}
Equation~(\ref{tildejcorr}) then gives
\begin{equation}
\cK_{\tilde{j}}(\bq) = \kappa\bq^2 - \kappa^2 \bq^2 \cK_J(\bq) ~.
\end{equation}
Since the leading non-analyticity in $\cK_J(\bq)$ is $|\bq|$ (corresponding to $J_\mu$'s scaling dimension $2$), we conclude that the leading non-analyticity in $\cK_{\tilde{j}}(\bq)$ is $|\bq|^3$, as claimed earlier.
When considering this matter-gauge system, it is convenient to introduce an ``irreducible'' current correlation function \cite{MurthyShankarRMP, Herzog2007, Geraedts_rangedloops} via
\begin{equation}
\cK_{\tilde{j}}^\text{irred}(\bq) \equiv \frac{\cK_{\tilde{j}}(\bq)}{1 - \frac{1}{\kappa\bq^2} \cK_{\tilde{j}}(\bq)} ~.
\end{equation}
Physically, while $\cK_{\tilde{j}}(\bq)$ describes the current response relative to the probing field $c_\mu$, $\cK_{\tilde{j}}^\text{irred}(\bq)$ describes the response relative to the total field $c_\mu + \langle a_\mu \rangle_{c_\mu}$ seen by the matter, where $\langle a_\mu \rangle_{c_\mu}$ is the induced non-zero average of the internal gauge field $a_\mu$ due to the external probing field $c_\mu$. It is easy to check that
\begin{equation}
\cK_{\tilde{j}}^\text{irred}(\bq) = \frac{\bq^2}{\cK_J(\bq)} - \kappa\bq^2 ~.
\end{equation}
Defining ``conductivity'' $\sigma_J$ corresponding to the $J_\mu$ currents via the small-momentum behavior of $\cK_J(\bq)$, we have
\begin{equation}
\cK_J(\bq) = \sigma_J |\bq| \quad\implies\quad
\cK_{\tilde{j}}^\text{irred}(\bq) = \frac{1}{\sigma_J} |\bq| 
\quad \textrm{for small $|\bq|$} ~.
\end{equation}
The last equation is often stated as dual ``conductivity'' being inverse of the direct conductivity. The difference in the treatments of the $J_\mu$ currents and $\tilde{j}_\mu$ currents is that the latter have long-range interactions while the former have only short-range interactions, which affects their scaling dimensions for the total correlation functions.

%%%%%%%%%%%%%%%%%%%%%%%%%%%%%%%%%%%%%%%%%%%%%%%%%%%%%%%%%%%%%%%%%%%%%%%%%
%%%%%%%%%%%%%%%%%%%%%%%%%%%%%%%%%%%%%%%%%%%%%%%%%%%%%%%%%%%%%%%%%%%%%%%%%
\section{Emergent particle-hole symmetry and single Dirac cone in a 2D electron gas}

A very important application of the duality is the derivation of Son's proposed field theory for the particle-hole-symmetric composite Fermi liquid in the lowest Landau level (LLL) at filling factor $\nu = 1/2$. The starting point for this derivation is the observation that adding a magnetic field to our single Dirac fermion breaks ${\cal T}$ but not ${\cal C}$, and that the problem of such a Dirac fermion with particle-hole symmetry in magnetic field maps to the 2D electron gas (2DEG) half-filled LLL problem with emergent particle-hole symmetry \cite{WangSenthil2015,MetlitskiVishwanath2015}. The degeneracies associated with the magnetic field (i.e., frustration of the kinetic energy) are ``resolved'' by going to the dual fermions, which are again Dirac fermions but now at finite density as permitted by the action of ${\cal C}$ on these dual fields.

Dirac fermions also arise from different direct approaches to the 2DEG, particularly in the physics between fillings $\nu = 0$ and $\nu = 1$. A single massless Dirac fermion arises at a transition between $\nu = 0$ and $\nu = 1$ states in the Haldane model \cite{Haldanemodel} and also at a plateau transition in the Chalker-Coddington model \cite{Chalker}. In both cases, however, these often-useful Dirac fermions are at \emph{zero} magnetic field, and thus they are not related to the particle-hole-symmetric Dirac fermions used in the derivation of the Son's theory above.

\begin{figure}[h]
\includegraphics[width=\columnwidth]{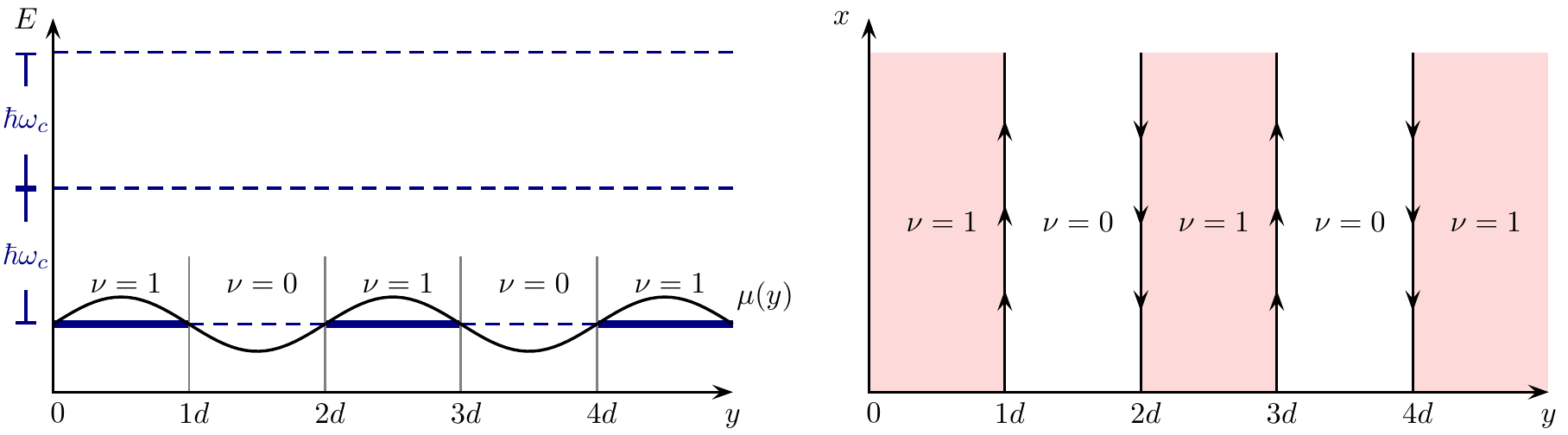}
\caption{Left: Electrons in a magnetic field form Landau levels separated by an energy $\hbar \omega_c$. A modulated chemical potential $\mu(y)$ produces regions alternating in space between filling factors $\nu=0$ and $\nu=1$. Right: The boundaries of these regions form an array of 1D electron `wires' of staggered chirality. In the $\hbar \omega_c \rightarrow \infty$ limit, hybridizing these modes yields a single massless Dirac cone in a magnetic field with an emergent particle-hole symmetry.}
\label{fig:2dd}
\end{figure} 

It is nevertheless possible to use our methods---in particular the quasi-1D construction---to derive Son's theory beginning from a 2DEG in magnetic field. To explicitly realize the Hamiltonian of Eqs.~(3,4) from the main text in this setting, consider electrons moving in a vector potential $A_\mu$ where $\partial_1 A_2 - \partial_2 A_1 = B$ and $A_0 = \mu + \delta \mu \sin(\frac{\pi}{d}y)$ with the Landau level separation $\hbar \omega_c \gg \delta \mu$ and magnetic length $\ell_B \ll d$. When $\mu$ is tuned such that the electrons are at half-filling, we obtain alternating $\nu=0$ and $\nu=1$ quantum Hall strips as shown in Fig.~\ref{fig:2dd}. The system therefore hosts an array of staggered-chirality edge modes precisely as utilized earlier. Let $\psi_y$ denote the edge field at interface $y$ (to better match our earlier notation we now set $d = 1$); odd and even $y$ respectively correspond to right- and left-movers. In the limit $\hbar \omega_c/\delta \mu \rightarrow \infty$, an exact symmetry emerges, 
\begin{equation}
 {\cal C} \psi_y {\cal C}^{-1} = \psi^\dagger_{y+1} ~,
\end{equation}
corresponding to the LLL particle-hole transformation---which interchanges $\nu = 0$ and $\nu = 1$ fillings---composed with a translation. Importantly, ${\cal C}$ is \emph{antiunitary}, which can be understood due to the generally complex LLL wavefunctions (see, e.g., \cite{Geraedts}). Because of the antiunitarity, the most general ${\cal C}$-symmetric, charge-conserving single-particle Hamiltonian that hybridizes the edge modes describes Dirac electrons with zero chemical potential but with a finite a magnetic field, which we analyzed in the main text. We emphasize that the emergent symmetry implemented by ${\cal C}$ has the same status as the particle-hole symmetry in a spatially isotropic $\nu = 1/2$ system: it only emerges upon projection into the LLL.

\section{Single (massive) Dirac cone in magnetic field from coupled wires}
\begin{figure}[h]
\includegraphics[width=.8\columnwidth]{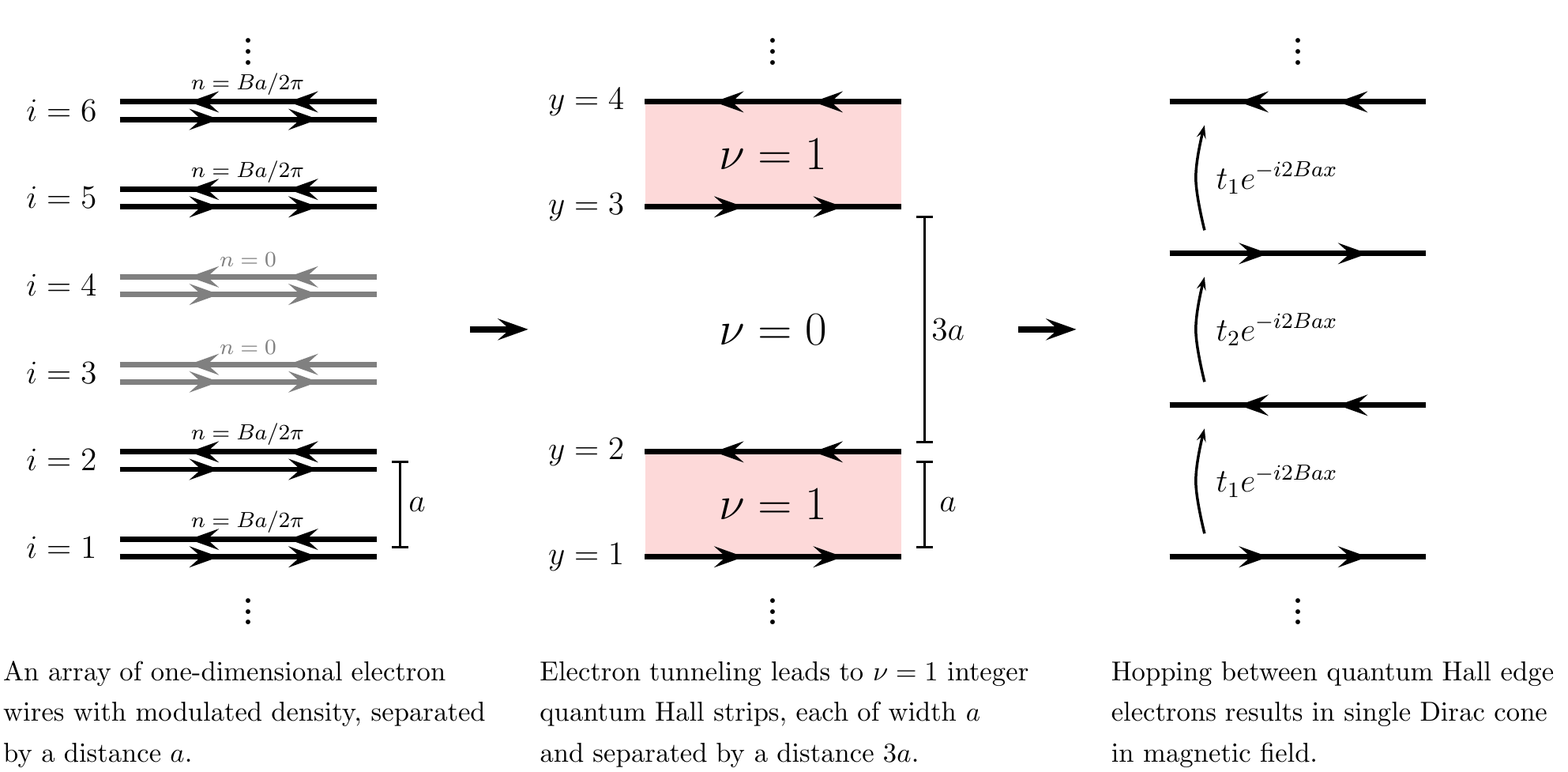}
\caption{A Dirac dispersion for electrons at $\nu=1/2$ can be obtained from strips of $\nu=1$ integer quantum Hall states. In this strict 2D construction there is no local symmetry that prohibits a mass term $m \sim t_1 - t_2$.}
\label{fig:2d}
\end{figure}

It is instructive to see how a (massive) Dirac cone in magnetic field arises more microscopically from a strictly 2D electron system. We therefore consider a two-dimensional system formed by an array of $N_\text{w}$ wires separated by a distance $a$. The wires are enumerated by integers $j$ and contain %a one-dimensional electron density $n_j$ 
(non-chiral) one-dimensional electron liquids at densities $n_j$. If the system is subjected to a perpendicular magnetic field of strength $B$, then the electrons are at half-filling when 
\begin{align}
 \nu \equiv \frac{2 \pi}{B N_\text{w} a}\sum_j n_j = \frac{1}{2} ~.
\end{align}
For example, this filling is realized when $M+1$ consecutive wires with density $n = \frac{B a}{2 \pi}$ alternate with $M+1$ wires with $n=0$ (shown in Fig.~\ref{fig:2d} for $M=1$). 
To analyze tunneling between the gapless edges of these strips, it is convenient to adopt the gauge $A_x = -B a j, A_y = 0$, and enumerate the edge states by integers $y$, where odd and even integers denote the bottom and top edges, respectively. 
Electrons in the $j$-th wire with non-zero density $n_j = n$ expanded in long-wavelength right- and left-moving fields are expressed as
\begin{equation}
c_j(x) \sim c_{j,R} e^{i (\pi n - B a j) x} + c_{j,L} e^{i (-\pi n - B a j) x}~.
\end{equation} 
Hopping of electrons between neighboring wires with $n_j \neq 0$ leads to $\nu = 1$ integer quantum Hall strips of width $Ma$ separated by a distance $(M+2)a$ \cite{KaneWires}.
We denote the surviving electronic edge modes of such a $\nu = 1$ region formed by wires $j, j+1, \dots, j+M$ as $c_{jR}, c_{j+M, L} \to \psi_{2y+1}, \psi_{2y+2}$. 
Tunneling between neighboring edges is then described by
\begin{align}
&h_1 = t_1 \psi_{2y+1}^\dagger \psi_{2y+2} e^{-i 2 \pi n x - i M a B x} + \text{H.c.} = t_1 \psi_{2y+1}^\dagger \psi_{2y+2} e^{-i (M+1) a B x} + \text{H.c.} ~,\\
&h_2 = t_2 \psi_{2y+2}^\dagger \psi_{2y+3} e^{i 2 \pi n x - i (M+2) a B x} + \text{H.c.} = t_2 \psi_{2y+2}^\dagger \psi_{2y+3} e^{-i (M+1) a B x} + \text{H.c.} ~,
\end{align}
where $h_1$ hops electrons across a given $\nu = 1$ strip with amplitude $t_1$, while $h_2$ hops electrons between adjacent $\nu = 1$ strips with amplitude $t_2$. Importantly, both kinds of hopping carry identical $x$-dependent phase factors, which correspond to an orbital magnetic field acting on the $\psi$ fermions. For $t_1 = t_2$ we therefore obtain precisely a single massless Dirac cone in a uniform magnetic field set by $B$. (We now see explicitly that the difference from the Chalker-Coddington network model \cite{Chalker}, where the plateau transition yields a Dirac theory at \emph{zero} magnetic field, arises because of kinematic restrictions imposed by the external magnetic field in our setup.) Within the present setup, however, $t_1 = t_2$ is not enforced by any microscopic symmetry---contrary to our surface construction---and requires fine-tuning. This is expected, since the exact particle-hole symmetry of the half-filled Landau level is not a local symmetry in any strictly two-dimensional system and requires projection into the lowest Landau level.

%%%%%%%%%%%%%%%%%%%%%%%%%%%%%%%%%%%%%%%%%%%%%%%%%%%%%%%%%%%%%%%%%%%%%%%%
%%%%%%%%%%%%%%%%%%%%%%%%%%%%%%%%%%%%%%%%%%%%%%%%%%%%%%%%%%%%%%%%%%%%%%%%
\section{Discrete symmetries for bosonic TI surface/$N=2$ QED$_3$ dual theories}
The action for $N=2$ QED$_3$ introduced in the main text exhibits the same discrete symmetries ${\cal T},{\cal C},{\cal I},{\cal M}$ as a free Dirac cone. In addition there is a local $\mathbb{Z}_2$ symmetry that interchanges the two fermion species,
\begin{align}
& {\cal R} \tilde \psi_{\sigma,y}(x) {\cal R}^{-1} = \tilde\psi_{-\sigma,y}(x) ~. \hspace{3cm}
\end{align}
Table \ref{tab.symm2} enumerates how these symmetries act on the various operators involved in the duality mapping. When considering specific bosonic TIs, one typically enforces only a subset of these symmetries. This dictionary may then be used to translate the symmetries of interest on the bosonic TI surface to symmetries in the dual gauge theory.

\begin{table}[h]
\caption{Action of discrete symmetries for bosonic TI surface/$N=2$ QED$_3$ duality (omitting phase factors).}
\begin{tabular}{| c| c| c| c | c| c|c|}
\hline
 & $\tilde \psi_{\sigma,y}(x)$& $\tilde \phi_{\sigma,y}(x)$& $\tilde \phi_{n,y}(x)$ &$\tilde \phi_{c,y}(x)$& $\phi_{c,y}(x)$& $ b_{\sigma,y}(x)$\\ \hline
 ${\cal T} \ldots{\cal T}^{-1} $ &$\tilde \psi_{\sigma,y+1}(x)$ & $-\tilde \phi_{\sigma,y+1}(x)$&$-\tilde \phi_{n,y+1}(x)$ 
 &$-\tilde \phi_{c,y+1}(x)$& $\phi_{c,y+1}(x)$ & $ b_{-\sigma,y+1}^\dagger(x)$\\
 \hline
 ${\cal C} \ldots{\cal C}^{-1}$ &$\tilde \psi^\dagger_{\sigma,y+1}(x)$ &$\tilde \phi_{\sigma,y+1}(x)$ &$\tilde \phi_{n,y+1}(x)$
 &$\tilde \phi_{c,y+1}(x)$& $-\phi_{c,y+1}(x)$ & $ b_{-\sigma,y+1}(x)$\\
 \hline
 ${\cal I} \ldots{\cal I}^{-1}$ &$\tilde \psi_{\sigma,-y+1}(-x)$ & $\tilde \phi_{\sigma,-y+1}(-x)$&$\tilde \phi_{n,-y+1}(-x)$
 &$\tilde \phi_{c,-y+1}(-x)$& $\phi_{c,-y+1}(-x)$ &$ b_{\sigma,-y+1}(-x)$\\
 \hline
 ${\cal M} \ldots{\cal M}^{-1}$ &$\tilde \psi_{\sigma,-y}(x)$ &$\tilde \phi_{\sigma,-y}(x)$ & $\tilde \phi_{n,-y}(x)$
 &$\tilde \phi_{c,-y}(x)$& $-\phi_{c,-y}(x)$& $ b_{-\sigma,-y}^\dagger(x)$\\
 \hline
 ${\cal R} \ldots{\cal R}^{-1}$ &$\tilde \psi_{-\sigma,y}(x)$ &$\tilde \phi_{-\sigma,y}(x)$ &$-\tilde \phi_{n,y}(x)$
 &$\tilde \phi_{c,y}(x)$& $\phi_{c,y}(x)$& $ b_{-\sigma,y}(x)$\\
 \hline
\end{tabular}
\label{tab.symm2}\vspace{.5cm}
\end{table}

\section{Klein factors for $N=2$ QED$_3$ duality}
For the duality between the fermionic $N=2$ QED$_3$ and bosonic TI surface theory, it is natural to ask how the Klein factors present in the former are manifested in the latter. We show that the non-local duality transformation consistently maps between the two representations even at the level of exchange statistics. Rather than employing Klein factors $\eta_y$ as above, here it is convenient to equivalently ensure anticommutation between different fermion species by taking
\begin{align}
[\tilde \phi_{\sigma,y}(x),\tilde\phi_{y',\sigma'}(x') ]= i\,\pi(-1)^y \delta_{y,y'}[\delta_{\sigma,\sigma'} \text{sgn}(x-x') + \epsilon_{\sigma,\sigma'}]+ i\,\pi (1 - \delta_{y,y'})\sgn(y-y'),
\end{align}
where $\epsilon_{\sigma,\sigma'}$ is antisymmetric with $\epsilon_{+-}=1$. Charge and neutral modes $\tilde \phi_{c/n} =( \tilde \phi_+ \pm \tilde \phi_-)/2$ then obey
\begin{align}
&[\tilde\phi_{c,y}(x),\tilde\phi_{c,y'}(x') ]=i\,\frac{\pi}{2}(-1)^y\delta_{y,y'}\text{sgn}(x-x') +i\,\pi (1 - \delta_{y,y'})\text{sgn}(y-y')\\
&[\tilde\phi_{n,y}(x),\tilde\phi_{n,y'}(x') ]=i\,\frac{\pi}{2}(-1)^y \delta_{y,y'}\text{sgn}(x-x')\\
&[\tilde\phi_{c,y}(x),\tilde\phi_{n,y'}(x') ]=-i\,\frac{\pi}{2}(-1)^y\delta_{y,y'} ~,
\end{align}
while fields dual to $\tilde \phi_{c,y}$ satisfy
\begin{align}
&[\phi_{c,y}(x),\phi_{c,y'}(x') ]=-i\,\frac{\pi}{2}(-1)^y\delta_{y,y'}\text{sgn}(x-x') -i\,\pi (1 - \delta_{y,y'})\text{sgn}(y-y')~.
\end{align}
Finally, the phases of dual bosons $\varphi_{\sigma,y}=(\phi_{c,y} +\sigma \tilde\phi_{n,y})$ exhibit commutators
\begin{align}
&[\varphi_{\sigma,y}(x),\varphi_{\sigma,y'}(x')]
=0 \ \ \text{mod} \ \ \ 2\, i\,\pi \\
&[\varphi_{\sigma,y}(x),\varphi_{-\sigma,y'}(x')]
=- i\,\pi(-1)^y \delta_{y,y'}\text{sgn}(x-x') -
i\,\pi (1 - \delta_{y,y'})
\end{align}
Note that the mutual phase of $\pi$ in the last line does not affect the interpretation of $b_\sigma \sim e^{i\varphi_{\sigma,y}}$ as bosons. Since bosons of species $\sigma =+$ and $\sigma = -$ are distinguishable, only a full braid (rather than a single exchange) is meaningful; there the acquired phase is an integer multiple of $2\pi$ as required for bosons.

\end{document}